\documentclass[11pt]{article}
\usepackage{pdfsync}
\usepackage{graphicx} 
\usepackage{subfig}
\usepackage{color}
\usepackage{amsfonts}
\usepackage{amssymb}
\usepackage{epsfig}
\usepackage{dcolumn}
\advance \textheight by \topskip
\usepackage{amsmath}
\usepackage[english]{babel}
\makeatletter
 \@addtoreset{equation}{section}
\makeatother

\usepackage{amsmath,amssymb}
\makeatletter \@addtoreset{equation}{section}
\makeatother
\def\be{\begin{equation}}

\def\be{\begin{equation}}
\def\ee{\end{equation}}

\def\cP{{\mathcal{P}}}

\def\cD{{{\mathcal{D}}}}
\def\cL{{{\mathcal{L}}}}

\def\A{\mathbb A}

\def\Z{\mathbb Z}
\def\R{\mathbb R}

\usepackage{amsmath,amssymb}
\def\bea{\begin{eqnarray}}
\def\eea{\end{eqnarray}}
\def\barray{\begin{array}}
\def\earray{\end{array}}

\def\sn{\mathrm{sn}}
\def\cn{\mathrm{cn}}
\def\dn{\mathrm{dn}}

\def\sech{\mathrm{sech}}

\def\cP{\mathcal{P}}
\def\cL{\mathcal{L}}
\def\cS{\mathcal{S}}
\def\cQ{\mathcal{Q}}
\def\cD{\mathcal{D}}
\def\cV{\mathcal{V}}

\begin{document}

\title{
{\bf Chiral asymmetry in propagation of soliton defects in crystalline backgrounds
}}

\author{Adri\'an Arancibia and Mikhail S. Plyushchay  \\
[4pt]
{\small \textit{
Departamento de F\'{\i}sica,
Universidad de Santiago de Chile, Casilla 307, Santiago 2,
Chile  }}\\
 \sl{\small{E-mails: 
adaran.phi@gmail.com, mikhail.plyushchay@usach.cl
}}
}
\date{}
\maketitle

\begin{abstract}

By applying Darboux-Crum  transformations
to 
a
 Lax pair formulation of the Korteweg-de Vries (KdV) equation,
we construct new sets of multi-soliton solutions to it as well as to 
the modified Korteweg-de Vries (mKdV) 
equation.  
The obtained solutions  
exhibit a chiral asymmetry in propagation 
of different types 
of
defects in crystalline backgrounds.
We show that the KdV 
solitons of pulse and compression modulation types,  
which support
bound states 
in,
respectively,   
semi-infinite  and  finite 
forbidden 
bands
 in the spectrum of the perturbed quantum
one-gap
Lam\'e system,  propagate in opposite directions with respect
to the asymptotically periodic  background.
A similar but more complicated picture also appears for 
multi-kink-antikink 
mKdV  solitons  that  propagate 
with a privileged direction
over 
the
 topologically 
trivial 
or topologically nontrivial  
crystalline background 
in dependence on position of energy levels of  
trapped bound states  in  
spectral gaps of the associated Dirac system.  
Exotic
$N=4$ nonlinear supersymmetric structure  incorporating Lax-Novikov integrals 
of a pair of perturbed Lam\'e systems is shown to underlie
the Miura-Darboux-Crum construction.
It  
unifies
the KdV and mKdV solutions,
detects the
defects and distinguishes their types, and
identifies the types
of crystalline  backgrounds. 
\end{abstract}

\vskip.5cm\noindent
\section{Introduction}\label{Introsec}

Nonlinear integrable systems play  an important role 
in a variety of areas of physics and its applications
\cite{Lamb,NMPZ,Heeger,Draz,Kivshar,Kundu,Malomed,Kuznetzov}.
A Lax pair formulation in particular cases of the 
Korteweg-de Vries (KdV) and modified Korteweg-de Vries  (mKdV)
equations  relates their soliton, kink, kink-antikink,
and  crystalline  type solutions with 
reflectionless and finite-gap systems 
of quantum mechanics 
\cite{KayMos,Matveev,AlgGeo}. 
This
formulation  for the KdV and mKdV systems 
includes, respectively,   the 
stationary  
Schr\"odinger   
and
Dirac equations (the latter with a scalar potential)
as the eigenstate equation.

The inverse scattering method allows to construct soliton solutions for the KdV equation
in  analytical form for 
the
asymptotically free, homogeneous
boundary  conditions.
On the other hand, 
such analytic solutions can also be constructed 
by exploiting  the covariance of the Lax pair 
under Darboux transformations and their generalization in the form of 
Darboux-Crum transformations. {}From the spectral point of view, 
these transformations correspond to
adding  soliton defects by means of introducing  
bound states into the spectrum of the initial potential
\cite{Matveev}.

Recently in \cite{5A}, by employing Darboux-Crum transformations,
we constructed a generalization of 
reflectionless potentials 
by introducing soliton defects into 
the
periodic (crystalline) background 
of the one-gap Lam\'e system
\be
    U_{\text{Lam\'e}}(x)=2 k^2 
    \text{sn}^2
	\left(
	x \left|k\right.\right)+const
	 \,.
\ee
As a result, we obtained one-gap potentials for  the Schr\"odinger system with 
an
arbitrary number of bound states in
the
lower and intermediate forbidden bands, which are trapped
by the 
soliton defects.

Using a relation between Darboux transformations and supersymmetric quantum mechanics,
it was also  possible to construct 
finite-gap Dirac (Bogoliubov-de Gennes) Hamiltonian operators,  
in which scalar potentials carry 
perturbations 
of the kink and kink-antikink type  introduced into the 
crystalline backgrounds 
of different  nature. 

One such periodic background
corresponds to the superpotential of the form 
\cite{BM,DF,CJNP,PAN}
\be\label{VKCGN}
	V^{KC}(x)=\frac{d}{dx}\log {\text{dn}\left(
	x  \left|k\right.\right)}=
	 -
	 k^2 \frac{ \text{cn}\left(
	 x  
	\left|k\right.\right) \text{sn}\left( 
	x  \left|k\right.\right)}	
	{\text{dn}\left(
	 x  \left|k\right.\right)}\,.
\ee
This finite-gap  Dirac scalar potential 
 appeared as a  solution in 
the Gross-Neveu model 
with a discrete chiral symmetry \cite{Th1,BD1}, 
and was identified there as the kink crystal. 
In the  QCD framework, the kink crystal 
can be considered as 
the phase corresponding  to the 
crystalline color superconductor~\cite{CCS},
that  is the QCD analogue of the 
Larkin-Ovchinnikov-Fulde-Ferrell
phase in the context of electron 
superconductivity~\cite{LO,FF}.
The  kink crystal type solution  (\ref{VKCGN})
also  found some applications  
in the physics of conducting polymers \cite{SaxBis,BraGor}.
In the construction from  \cite{5A},
the Dirac Hamiltonian operator 
with the perturbed  scalar potential  
of the form (\ref{VKCGN}) possesses  a spectrum
having  a central allowed band between the conduction band and 
the Dirac sea,  
and the bound states appear there symmetrically in both forbidden bands
(gaps). 

There is yet  another possibility for asymptotic behaviour  in the form of the 
kink-antikink, $V^{KA}$,  (or, antikink-kink, $V^{AK}$) crystal \cite{PAN}, 
\be\label{VKAAK}
	V^{KA}(x)=
	{\rm Z}( 
	x+\beta\left|k\right. )-
	{\rm Z}(
	x\left|k\right.)-{\rm z}(\beta\left|k
	\right.)
	\,,\qquad V^{AK}(x)=-V^{KA}(x)\,,
\ee
which was found as a solution in the Gross-Neveu model with a bare mass term \cite{Th2}.
It allowed to obtain scalar potentials with two  
 permitted bands between the 
Dirac sea and conduction band, 
with arbitrary number of  bound states in central and/or other two forbidden bands. 
Here $  {\rm z}\left(\beta\left|k\right. \right)$ is a  constant
given in terms of the  Jacobi Zeta and elliptic functions,
$  {\rm z}\left(\beta\left|k\right. \right)=
{\rm Z}\left(\beta+i{\rm {\bf K}}' \left|k\right.\right)+
i\frac{\pi}{2{\rm {\bf K}}}={\rm Z}\left(\beta \left|k\right.\right)+
\cn\,
(\beta \left|k\right.) \,\dn\,(\beta \left|k\right.)/\sn\,(\beta \left|k\right.)\,$.
An additional  change of the global topology of the 
Dirac scalar potential 
provided us 
with
reflectionless kink-antikink perturbations over the 
kink-type 
crystalline
background.

 A natural question we address in this paper is the following:  
how the dependence on  time can be
`reconstructed' for
the finite-gap 
 Schr\"odinger and 
Dirac crystalline  potentials with defects,
found  in \cite{5A},
so that 
they will be
 solutions 
for  the KdV and  mKdV  equations?
We refer here  to the evolution parameter 
in the corresponding  nonlinear integrable 
systems;
from the perspective of the
eigenvalue problems,  it
 is associated with the isospectral deformation
of the potentials.
Answering this question, we  reveal  a  phenomenon 
of a chiral asymmetry in propagation 
of different types 
of
 soliton defects in crystalline backgrounds, which,
as we believe,  could be interesting, particularly  
from the point of view of applications. 

The paper is organized as follows.  In the next 
Section \ref{Laxpairsec},
that  is of a very brief review character, we consider  the Lax pair
formulation for the KdV equation, discuss its Darboux covariance, 
and apply the Darboux and Darboux-Crum transformations
to obtain multi-soliton solutions for the KdV equation.
Section \ref{cnostat} is devoted to the construction 
of solutions to the auxiliary problem 
corresponding to  the Lax pair 
with the one-gap Lam\'e potential 
taken as the stationary, periodic  cnoidal solution 
to the KdV equation. 
On the basis of the solutions to the auxiliary problem, 
in  Section \ref{MulKdVcryst} 
we construct  solutions to the KdV equation
by employing Darboux-Crum transformations.
In this way we obtain  the KdV solutions  
that describe propagation of solitons of 
the  potential well (pulse)  and compression modulation types
in the background  of asymptotically periodic 
cnoidal wave. We also discuss there the issue of
velocities of the defects. 
In Section \ref{KdVmKdVuni}, we first show how the  Miura and 
Darboux-Crum transformations  together with 
the Galilean symmetry of the KdV equation can be employed 
for the construction of solutions to the mKdV equation 
on the basis of the KdV solutions.
Then we discuss  the exotic $N=4$ nonlinear supersymmetric 
structure  incorporating Lax-Novikov integrals 
of a pair of perturbed Lam\'e systems.
This supersymmetry 
underlies the Miura-Darboux-Crum construction and unifies  
the KdV and mKdV solutions.
In Section \ref{mKdVsolSec}  we first discuss briefly  the
 asymptotically free  mKdV  solutions 
corresponding to the multi-kink-antikink solitons propagating over the
kink or kink-antikink background.
Then we  construct  a  much more rich set 
of multi-kink-antikink 
soliton solutions  for the mKdV equation  
which propagate over
the
 topologically 
trivial  or topologically nontrivial 
crystalline backgrounds.
The last Section \ref{DiscOutlook} 
is devoted to the discussion and outlook.

\section{Lax pair for the
 KdV equation, Darboux transformations,
 and multi-soliton solutions }\label{Laxpairsec}
 
\subsection{Auxiliary spectral problem  for the KdV equation}\label{Laxpairsec1}
Consider a linear  system
\be 
\label{01}
	L\phi=\lambda\phi\,,\qquad
	\frac{\partial \phi}{\partial t}=P\phi\,
\ee 
for a function $\phi=\phi(x,t,\lambda)$.
It is assumed that $L$ and $P$ are some  
(in general, matrix) differential operators in space coordinate $x\in \R$
with  coefficients that can  also 
depend on evolution parameter $t$.
If the evolution in $t$ generated  by $P$ is isospectral, ${d\lambda}/{dt}=0$,
then the condition of consistency for the system  (\ref{01}) reduces to 
the Lax equation
\be\label{LtPL}
\frac{\partial L}{\partial t}=[P,L]\,.
\ee
For the choice of the Lax pair in the form of differential operators
\bea\label{LP1}
	L&=&-\partial^2_x+u\,,\\
	P&=&
	-4\partial^3_x+6u\partial_x+3u_x\,, \label{LP2}
\eea
equation (\ref{LtPL}) reduces to the 
Korteweg-de Vries equation for the scalar field $u=u(x,t)$, 
\bea\label{KdV}
	u_t=6uu_x-u_{xxx}\,.
\eea

\subsection{Darboux covariance  of the  KdV equation}\label{Laxpairsec2}

The system of equations
corresponding to  the Lax  pair (\ref{LP1}), (\ref{LP2}),
\bea\label{LaxKdV}
	(-\partial^2_x+u)\Psi(x,t,\lambda)&=&\lambda\Psi(x,t,\lambda)\,,\\
	\frac{\partial}{\partial t}\Psi(x,t,\lambda)&=&(-4\partial^3_x+6u\partial_x+
	3u_x)\Psi(x,t,\lambda)\,,\label{LaxKdV1}
\eea
is covariant under Darboux transformations \cite{Matveev}
\bea
	u(x,t) &\rightarrow& {u}_1(x,t)=u(x,t)-2(\log \Psi(x,t,\lambda_1))_{xx}\,,\label{D1}\\
	\Psi(x,t,\lambda)&\rightarrow&{\Psi}_1(x,t,\lambda)=
	\frac{W(\Psi(x,t,\lambda_1),\Psi(x,t,\lambda))}{\Psi(x,t,\lambda_1)}\,,
	\label{D2}
\eea
where $W$ is the Wronskian, $W(f,g)=fg_{x}-f_{x}g$.
This follows from the observation that if
Eqs. (\ref{LaxKdV}) and (\ref{LaxKdV1})
are fulfilled for 
 $\Psi(x,t,\lambda)$,
 and   $\Psi(x,t,\lambda_1)$ satisfies the same equations 
 with $\lambda$ changed
 for $\lambda_1$, then  
 ${\Psi}_1(x,t,\lambda)$
 obeys the equations (\ref{LaxKdV}), (\ref{LaxKdV1})
 with $u(x,t)$ changed for $u_1(x,t)$.
 As a  consequence,
 if $u(x,t)$ is a solution of the KdV equation
 (\ref{KdV}), then $u_1(x,t)$  
 obeys  the same equation. 
 
 This result can be extended for a 
 finite sequence of consecutive Darboux transformations, 
\bea\label{uxtn}
	u(x,t)&\rightarrow&{u}_m(x,t)=u(x,t)-2(\log W(\Psi(x,t,\lambda_1),\ldots,\Psi(x,t,\lambda_m)))_{xx}\,,\\
	\Psi(x,t,\lambda)&\rightarrow&{\Psi}_m(x,t,\lambda)=
	\frac{W(\Psi(x,t,\lambda_1),\ldots,\Psi(x,t,\lambda_m),\Psi(x,t,\lambda))}
	{W(\Psi(x,t,\lambda_1),\ldots,\Psi(x,t,\lambda_m))}\,,
\eea
that is known as the Darboux-Crum  transformation of order 
$m$ \cite{Matveev}.
For the sake of simplicity of notations, 
the  dependence  of  ${u}_m(x,t)$ as well as of 
$\Psi_m(x,t,\lambda)$
on  the 
 spectral parameters 
$\lambda_1,\ldots,\lambda_m$ is not shown explicitly.

One can define a first order differential operator 
$A_1$ in terms of the function 
$\Psi(x,t,\lambda_1)$
that underlies the Darboux transformation construction
(\ref{D1}), (\ref{D2}),
\be\label{A1def}
	A_1=\Psi(x,t,\lambda_1)\partial_x\frac{1}{\Psi(x,t,\lambda_1)}=
	\partial_x-V_1(x,t),\qquad V_1(x,t)=
	(\log \Psi(x,t,\lambda_1))_x\,.
\ee
The operator $A_1$ and its Hermitian conjugate, $A_1^\dagger$,  
intertwine the Schr\"odinger operators $L=-\partial_x^2+u(x,t)$
and $L_1=-\partial_x^2+u_1(x,t)$,
\be\label{A1inter}
 A_1L=L_1A_1,\qquad A^\dag_1L_1=LA^\dag_1,
\ee
and factorize them, 
\be\label{A1factor}
	A^\dag_1A_1=L-\lambda_1,
	\qquad A_1A^\dag_1=L_1-\lambda_1\,.
\ee
The two relations in (\ref{A1factor}) 
are equivalent to Riccati equations for the function
$V_1$, 
\be
V_1^2+V_{1x}=u-\lambda_1,
\qquad V_1^2-V_{1x}={u}_1-\lambda_1\,.
\ee

For the  successive Crum-Darboux transformations
of orders $m-1$ and $m$,
the corresponding Schr\"odinger  operators 
$L_{m-1}$ and  $L_m$ are related by means of 
the first order differential operator 
\be
	A_m=(\A_{m-1}{\Psi}(x,t,\lambda_m))
	\partial_x\frac{1}{(\A_{m-1}{\Psi}(x,t,\lambda_m))}=\partial_x-V_m(x,t),\qquad
	m=1,\dots\,,
\ee
and its conjugate,  
where 
$\A_n=A_n A_{n-1}\ldots A_1$, $A_0\equiv 1$. 
Then, as a generalization of  (\ref{A1inter}) and (\ref{A1factor}), we have 
$ A_m L_{m-1}=L_mA_m,\qquad A^\dag_m L_m=L_{m-1}A^\dag_m,$
and $A^\dag_m A_m=L_{m-1}-\lambda_m,
	\qquad A_mA^\dag_m=L_m-\lambda_m$,
where  $L_0\equiv L$.

Superpotential $V_m$ can be presented in terms of the 
Wronskians \cite{AGPfer}, 
\be\label{vdef}
	V_m(x,t)=\Omega_{m-1}(x,t)-\Omega_m(x,t)\,,\qquad
	\Omega_m=-\left(\log W(\Psi(x,t,\lambda_1),
	\ldots,\Psi(x,t,\lambda_m))\right)_x\,,
\ee
where we assume $\Omega_0=0$ and $W(\Psi(x,t,\lambda_1))=\Psi(x,t,\lambda_1)$.
It satisfies the Riccati equations
\be\label{vmlam}
	V_m^2(x,t)+(V_{m}(x,t))_x=u_{m-1}(x,t)-\lambda_m,
	\qquad V^2_m(x,t)-
	(V_{m}(x,t))_x={u}_m(x,t)-\lambda_m\,.
\ee

\subsection{Multi-soliton solutions of the KdV equation}\label{multisolKdV}

The described picture with Darboux-Crum  
transformations  can be illustrated by the  
construction of  multi-soliton solutions for  the KdV equation.

\begin{figure}[h]
	\centering
	\includegraphics[scale=3]{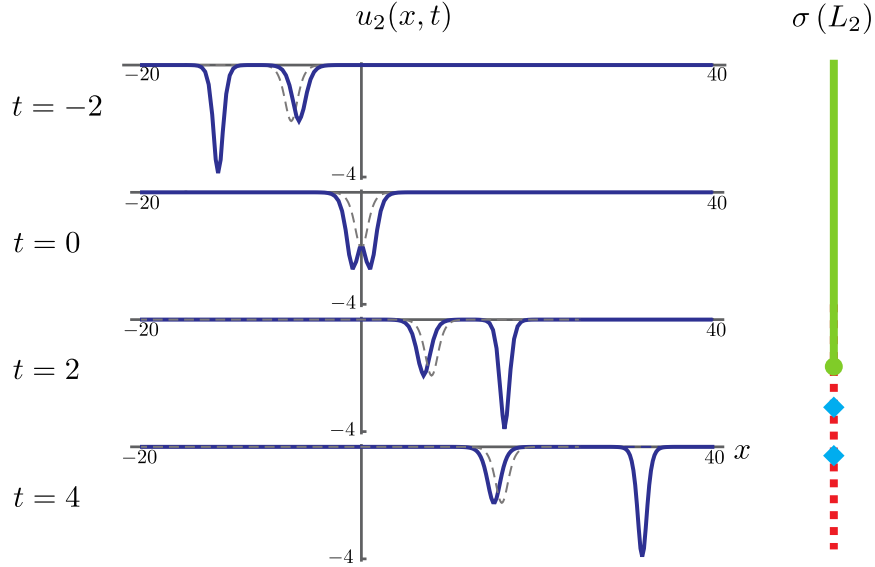}\\
	\caption{The KdV  two-soliton solution with 
	$\kappa_1 = 1$, $\kappa_2 = 1.4$, and
	$x_{0i} = 0$, $i = 1, 2$, is shown by a continuous line.
	 Propagation of a 
	one-soliton solution 
	with $\kappa_1=1$ and $x_{01}=0$ is depicted by a dashed line.  
	Initial phases are chosen so that 
	the two-soliton and one soliton solutions at $t=0$ are symmetric 
	with respect to the point $x=0$.
	On the right, the spectrum  of the associated two-soliton Schr\"odinger 
	operator $L_2$ is shown. The continuous green line corresponds 
	to a doubly degenerate semi-infinite continuos part of the spectrum
	with eigenstates $\psi^{\pm\kappa}(x,t)=\A_2e^{\pm iX^+(x,t;\kappa,x_0)}$, 
	while the filled circle indicates  a non-degenerate edge-state 
	described by the eigenfunction $\psi_{0}(x,t)=\A_2 1$.
	Dashed red line corresponds 
	to  non-physical semi-infinite part of the spectrum,
	inside which blue squares indicate energies 
	of the two bound states trapped by solitons and 
	described by eigenfunctions $\psi_1(x,t)=\A_2\sinh X^-_1$ and $
	\psi_2(x,t)=\A_2\cosh X^-_2$.}\label{figure1}
\end{figure}

The trivial solution of the KdV equation  is $u_0=0$. In this case 
$L_0=-\frac{\partial^2}{\partial x^2}$ corresponds to 
the Schr\"odinger operator for a  free particle system,
 and the evolution 
operator (\ref{LP2}) reduces to  $P_0=-4\frac{\partial^3}{\partial x^3}$.
The system (\ref{01})  
takes 
then
 the form
\be
	-\frac{\partial^2\Psi}{\partial x^2}=\lambda\Psi\,,
	\qquad
	\frac{\partial \Psi}{\partial t}=-4\frac{\partial^3\Psi}{\partial x^3}\,.\label{free2}
\ee
Acting on both sides of the first equation  by
$4\partial_x$ and summing up  the result
with the second equation, we obtain $\partial_t\Psi=4\lambda\partial_x\Psi$. 
Therefore,  $\Psi(x,t,\lambda)=\Psi(x+4\lambda t,\lambda)$.
For $\lambda=0$, $\lambda=-\kappa^2<0$, and $\lambda=\kappa^2>0$,
the pairs of linearly 
independent solutions of the system (\ref{free2}) can be chosen in the form
\be\label{Pf0}
\Psi(x,t,\lambda=0)=\{1,\, x\}\,,
\ee
\be\label{Pf1}
 \Psi(x,t,\lambda=-\kappa^2)=
\{\cosh X^-,\, \sinh X^-\},\qquad 
\Psi(x,t,\lambda=\kappa^2)=\{\cos X^+,\,\, \sin X^+\}\,,
\ee
where $X^\mp= \kappa(x-x_0\mp 4\kappa^2t)$.
To apply the Darboux-Crum  transformation to a trivial solution $u_0=0$,
we have different possibilities for choosing wave functions from the sets 
(\ref{Pf0}), (\ref{Pf1}).
The choice $\Psi(x,t,0)=x$ gives rise by means of the first order Darboux transformation
(\ref{D1}) to the simplest but singular, time-independent solution 
of the KdV equation, $u_1(x)=2/x^2$. 
The \emph{non-singular} solutions for the  KdV equation are generated by
choosing appropriately the eigenstates with eigenvalues $\lambda<0$, 
\be\label{sol_n}
 {u}_{n}(x,t)=-2\frac{\partial^2}{\partial x^2}
 \log W(\cosh X^-_1,\sinh X^-_2\ldots,
 f(X^-_{n}))\,,\qquad
 X^-_j=\kappa_j(x-x_{0j}-4\kappa^2_jt)\,,
 \ee
where the last argument in the Wronskian is 
$f(X^-_{n})=\sinh X^-_n$ if $n$ is even,  $n=2l$,
and 
 $f(X^-_{n})=\cosh X^-_n$ for odd $n=2l+1$;
$x_{0j}$ are the translation (phase) parameters, and 
the scale parameters $\kappa_j$ have to  obey the 
inequalities
$0<\kappa_1<\kappa_2<\kappa_3<\ldots<\kappa_n$.
Functions (\ref{sol_n}) correspond to 
$n$-soliton solutions of the KdV  equation.
The case $n=2$ is illustrated in Fig. \ref{figure1}.
When solitons in the solution (\ref{sol_n}) are well separated,
the propagation to the right of the $j$-th soliton 
can be  characterized by  the 
speed $\mathcal{V}_j=4\kappa^2_j$ and amplitude $2\kappa^2_j$.

\section{Spectral problem with the  cnoidal background}\label{cnostat}

The simplest stationary periodic solution to the KdV equation 
(\ref{KdV}) can be presented in 
 a form 
\be\label{u0Lame}
	u(x)=u_{0,0}(x)=2 k^2 \mu^2 
	\text{sn}^2\left(\mu x\left|k\right.\right)-\frac{2}{3} \left(1+k^2\right) \mu^2\,,
\ee
where 
$\text{sn}\,(u\vert k)$ is the Jacobi elliptic function, 
whose real and imaginary periods depend 
on the modular parameter $0<k<1$,
$\mu>0$ is a free (scale) parameter, and the sense of the indices in $u_{0,0}$
will be clarified below.
Due to the $t$-independence of solution 
(\ref{u0Lame}), Lax equation
(\ref{LtPL}) reduces to the condition 
of commutativity of the corresponding operators (\ref{LP1}) 
and (\ref{LP2}) constructed on the basis of (\ref{u0Lame}),
\be\label{LP0}
	[L,P]=0\,.
\ee
Relation (\ref{LP0})  guarantees the existence of the common basis 
for the operators  $L$ and $P$. We then 
look for the solutions of the system of
equations (\ref{LaxKdV}), (\ref{LaxKdV1}) in the form 
\be\label{Psixt}
\Psi(x,t,\lambda)=\Phi(x,\alpha)\exp(\pi(\alpha)t)\,,
\ee
where $\Phi(x,\alpha)$ is a common eigenstate 
of $L$ and $P$, 
$L\Phi(x,\alpha)=\lambda(\alpha)\Phi(x,\alpha)$,
$P\Phi(x,\alpha)=\pi(\alpha)\Phi(x,\alpha)$.
The sought for state $\Phi(x,\alpha)$ is
\cite{5A,PAN,WW}
\be\label{Lla} 
	\Phi(x,\alpha)=
	\frac{{\rm H}\left(\mu x+\alpha \vert k\right) }{\Theta
	 \left(\mu x\vert k\right)}
	e^{ -\mu x {\rm Z}\left(\alpha \vert k 
	\right)},
\ee
where ${\rm H}$, $\Theta$ and ${\rm Z}$ 
are Jacobi's Eta, Theta and Zeta functions,
while the corresponding 
eigenvalues are 
\be\label{lama}
	\lambda(\alpha\vert k)=\mu^2\left(\dn^2
	\big(\alpha|k\right)-\frac{1}{3}(1+k'{}^2)\big)\,,
\ee
\be\label{pia}
	\pi(\alpha\vert k)=-4 k^2 \mu^3 \text{sn}
	\left(\alpha \left|k\right.\right) \text{cn}\left(\alpha \left|k\right.\right) 
	\text{dn}\left(\alpha \left|k\right.\right)\,.
\ee
Notice that $\lambda(-\alpha)=\lambda(\alpha)$, 
$\pi(-\alpha)=-\pi(\alpha)$,
$\pi(\alpha)=2\mu\frac{d \lambda(\alpha)}{d\alpha}$,
and 
\be\label{pi2lam}
	\pi^2(\alpha)=
	-16\big(\lambda(\alpha)-E_0\big)
	(\lambda(\alpha)-E_1)(\lambda(\alpha)-E_2)\,,
\ee
where 
\be\label{E0E1E2}
	E_0=-\frac{1}{3}(1+k'{}^2)\mu^2\,,\qquad 
	E_1=\frac{1}{3}(1-2k^2)\mu^2\,,\qquad
	E_2=\frac{1}{3}(1+k^2)\mu^2\,, 
\ee
and  
$k'=\sqrt{1-k^2}$ is 
the
complementary
modular parameter.
Relations (\ref{LP0}) and (\ref{pi2lam}) 
correspond to the Burchnall-Chaundy
theorem \cite{BurChau}, and 
reflect the one-gap nature of the potential 
(\ref{u0Lame}) \cite{NMPZ,AlgGeo}, see below.

Taking into account 
the double periodicity of the Jacobi $\sn$, $\cn$ and $\dn$ functions,
without loss of generality one can 
suppose  that a dimensionless parameter $\alpha$ takes values 
in the rectangle in the complex plane with  vertices in\footnote{
Here, ${\rm{\bf K}}={\rm{\bf K}}(k)$  
is a complete elliptic integral of the first kind and  
${\rm{\bf K}}'={\rm{\bf K}}(k')$.
}
$0$, ${\rm{\bf K}}$, ${\rm{\bf K}} + i {\rm{\bf K}}'$ and 
$i {\rm{\bf K}}'$. 
On the 
border 
of this \emph{fundamental 
$\alpha$-rectangle}, 
eigenvalues $\lambda(\alpha)$ of the Schr\"odinger 
operator $L$ take real values, 
while the eigenvalue $\pi(\alpha)$ 
of the Lax operator $P$ takes nonzero real values on the 
horizontal borders of this rectangle where
$\alpha=\beta^+$, $0<\beta^+<{\rm{\bf K}}$, and 
$\alpha=\beta^-+i{\rm{\bf K}}'$, $0<\beta^-<{\rm{\bf K}}$.
On the other hand, on  both vertical borders of the rectangle,
$\alpha=i\gamma^+$, $0\leq \gamma^+ < {\rm{\bf K}}'$, 
and $\alpha={\rm{\bf K}}+i\gamma^-$, $0\leq\gamma^-\leq {\rm{\bf K}}'$,
 $\pi(\alpha)$ takes pure imaginary values, 
 turning into zero at the vertices 
 $0$, ${\rm{\bf K}}$ and ${\rm{\bf K}} + i {\rm{\bf K}}'$. 
 The vertex  $i {\rm{\bf K}}'$ is the third order pole 
 of $\pi(\alpha)$ and the second order pole of 
 the function $\lambda(\alpha)$.
Moreover, corresponding eigenfunctions (\ref{Lla}) 
are unbounded  functions  on the
horizontal edges,  but they are bounded 
on the vertical 
borders of the rectangle. 
Summarizing, in correspondence with the specified  properties,
the solutions (\ref{Psixt}) of the auxiliary spectral problem 
for the KdV equation are bounded 
on the vertical borders  of the rectangle,
except the vertex point $\alpha=i {\rm{\bf K}}'$,
and they  are unbounded 
on the horizontal borders
except the vertices $\alpha=0,\, {\rm{\bf K}},\, 
{\rm{\bf K}}+i{\rm{\bf K}}'$.

The
Schr\"odinger operator $L$ with potential
(\ref{u0Lame}) describes quantum mechanical periodic one-gap Lam\'e system,
for which the upper horizontal line 
$\alpha=\beta^-+i{\rm{\bf K}}'$
corresponds to the semi-infinite forbidden band 
with energy values $-\infty<E(\alpha)=\lambda(\alpha)<E_0$,
and  the lower horizontal line $\alpha=\beta^+$ 
corresponds to  a spectral 
gap with $E_1<E(\alpha)<E_2$. 
The corresponding bounds $E_0$, $E_1$ and $E_2$, 
defined in (\ref{E0E1E2}), satisfy the relations
$E_0<0$, $E_2>0$, while $E_1$
 can take positive or negative values in 
dependence on the value of the modular  parameter $k$.
 The vertical borders of the rectangular correspond to 
the allowed valence, $\alpha={\rm{\bf K}}+i\gamma^-$, 
$E_0\leq E(\alpha)\leq E_1$,
and conduction, $\alpha=i\gamma^+$, 
$E_2 \leq E(\alpha)<\infty$, bands.
With taking into account relations $\lambda(-\alpha)=\lambda(\alpha)$, 
$\pi(-\alpha)=-\pi(\alpha)$, the common eigenstates of $L$ and $P$ 
with opposite values for $\pi(\alpha)$ are obtained
from the described solutions with $\alpha$ restricted to 
the borders of the  indicated rectangle by changing 
$\alpha \rightarrow -\alpha$.  

Let us stress  that the  solutions (\ref{Psixt})
to the system of equations (\ref{LaxKdV}) and  (\ref{LaxKdV1})
have a factorizable dependence on 
the
evolution variable $t$
due to 
the $t$-independence of the corresponding KdV solution (\ref{u0Lame})
and the associated with it  Lax pair (\ref{LP1}), (\ref{LP2}).

\section{Multi-soliton defects in the crystalline background}\label{MulKdVcryst}

One can construct  three types of Darboux-Crum transformations 
based on the solutions  (\ref{Psixt}) to the 
auxiliary problem 
for the 
KdV equation. 
They will provide  us with new soliton solutions of the KdV equation 
that propagate over the stationary crystalline background  (\ref{u0Lame}) 
by deforming it.  These  are:

 i) soliton defects of the potential well (pulse)  type;

ii) soliton  defects of the compression modulations nature;

iii) the mixed case in which both types of the solitons are present.

\noindent
Soliton defects of the type i) are generated by 
Darboux-Crum transformations based on unbounded (non-physical) 
states from the lower  semi-infinite forbidden band 
in the spectrum of the one-gap Lam\'e system.
The construction of soliton defects of the type ii) 
requires the use of states from the gap of the 
quantum Lam\'e system. The employment of both types of 
non-physical states from the spectrum of the Lam\'e system 
generates solutions corresponding to the mixed case iii). 
Below we describe these three types of solutions.

\subsection{ Potential well (pulse)  type solitons}\label{case i)}

If we choose  common eigenstates (\ref{Psixt}) of $L$ and $P$ 
in the form of real-valued functions, then the new solutions generated by
Darboux-Crum transformations (\ref{uxtn}) also will be real.
The eigenfunctions 
$\Phi(x,\alpha=\beta^-+i{\rm{\bf K}}')$, $\beta^-\in (0, {\rm{\bf K}})$,
 of $L$ in  (\ref{Lla}) correspond to non-physical (unbounded) states 
with negative energies  from the lower forbidden band of one-gap Lam\'e system.
They are real-valued functions
modulo a global phase  factor \cite{5A,PAN}.  
Omitting a phase factor, we obtain a common real-valued eigenstates 
of $L$ and $P$,
\be \label{Fxtb}
	F(x,t,\beta^-)=\frac{\Theta \left(\mu x+\beta^- \left|k\right. \right) 
	}{\Theta \left(\mu x \left|k\right. \right)}
	\exp\Big(
	{ -\mu x{\rm z}(\beta^- \left|k\right. )+\pi(\beta^-+i{\rm {\bf K}}'\vert k)  t} \Big) \,,
\ee
where 
\be\label{zetabeta-}
	{\rm z}\left(\beta^-\left|k\right. \right)=
	{\rm Z}\left(\beta^-+i{\rm {\bf K}}' \left|k\right.\right)+
	i\frac{\pi}{2{\rm {\bf K}}}={\rm Z}\left(\beta^- \left|k\right.\right)+
	\frac{\cn\,
	(\beta^- \left|k\right.) \,\dn\,(\beta^- \left|k\right.)}{\sn\,(\beta^- \left|k\right.)}>0\,,
\ee
\be\label{pibeta-}
	\pi(\beta^-+i{\rm {\bf K}}'\vert k)=
	4 \mu^3 \frac{ \text{cn}\left(\beta^-\left|k\right.\right) 
	\text{dn}\left(\beta^-\left|k\right.\right)}{\text{sn}^3\left(\beta^-
	\left|k\right.\right)}>0\,.
\ee
The superpotential 
(\ref{VKAAK}) is given in terms of (\ref{Fxtb})  by $V^{KA}(x)=(\log F(x,t,\beta))_x$
with $\beta=\beta^-$ and $\mu=1$, that reduces to (\ref{VKCGN})
in the limit case $\beta^-={\rm {\bf K}}$.
On the basis of the functions (\ref{Fxtb}), we construct 
Darboux-Crum transformation of the form 
\be\label{u0lxt}
	 {u}_{0,l}(x,t)=u_{0,0}(x)-2(\log W(\mathcal{F}_+(1),\mathcal{F}_-(2),...,
	 \mathcal{F}_{(-1)^{l+1}}(l)))_{xx}\,,
\ee
where 
\be\label{Fdefpmj}
	\mathcal{F}_\pm(j)=C^-_jF(x,t,\beta^-_j)\pm\frac{1}{C^-_j}F(x,t,-\beta^-_j)\,,
	\qquad j=1,\ldots,l,
\ee
are combinations of the two linear independent solutions
(\ref{Fxtb}). Minus index in parameter coefficients $C^-_j$
indicates that (\ref{Fxtb}) are real eigenstates 
from the \emph{lower} forbidden band of the one-gap Lam\'e system. 
The choice 
\be
	 {\rm {\bf K}}>\beta^-_1>\beta^-_2>...>\beta^-_l>0\,,
	 \qquad 0<C^-_j<\infty\,,
\ee
guarantees the non-singular nature of 
the solution ${u}_{0,l}(x,t)$ for the KdV equation \cite{5A}.
This solution describes $l$ solitons of the
potential well type that propagate \emph{to the right}
in the stationary periodic background by deforming it.
For large negative and positive values of $t$, 
pulses are well separated, and each corresponding potential 
well supports a bound state with negative energy given by
(\ref{lama}) with $\alpha=\alpha_j=\beta^-_j+i{\rm{\bf K}}'$.
A deeper potential well soliton defect in asymptotically periodic background 
propagates faster,  and supports
the bound state with lower energy,
see Section \ref{veloSec} below.
Figures \ref{figure2} and \ref{figure3} 
 show such travelling soliton defects 
for the simplest cases of $l=1$ and $l=2$.
\begin{figure}[h]
	\centering
	\includegraphics[scale=2.7]{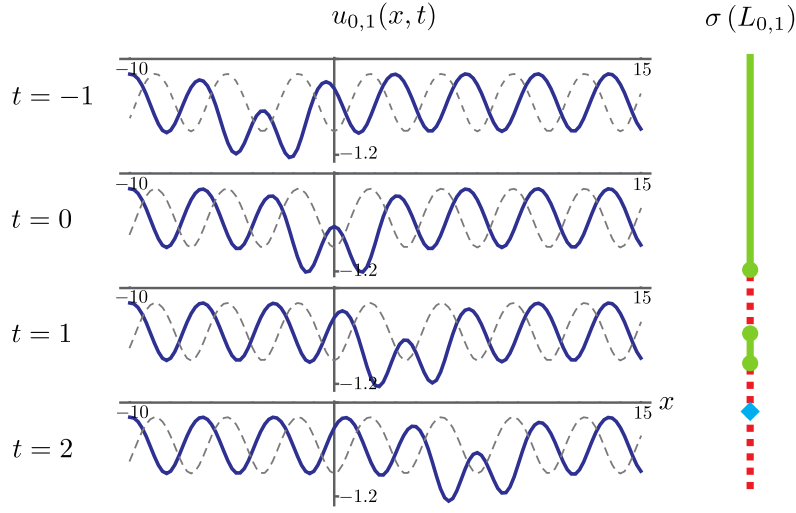}\\
	\caption{The KdV solution with 
	a
	 pulse soliton
	 propagating over
 the stationary crystalline background (\ref{u0Lame}) shown 
 by dashed line. 
 Here 
	 $\mu=1$, 
	$k=0.6$, $\beta^-_1=1.2$, $C^-_1=1$. Vertical line on the right 
	illustrates the spectrum of the perturbed  Lam\'e system 
	with lower and upper forbidden bands shown in dashed red.
	Filled semicircles indicate non-degenerate energy values at the 
	edges of the valence and conduction bands.
	 In the semi-infinite lower forbidden band, 
	there is a  bound state trapped by the soliton
	defect, that is shown by a blue square.  
	Parameters  in (\ref{Fdefpmj}) are chosen 
	so that the solution at $t=0$ is symmetric under the 
	space reflection
	$x\rightarrow -x$.	}\label{figure2}
\end{figure}
\begin{figure}[h]
	\centering
	\includegraphics[scale=3]{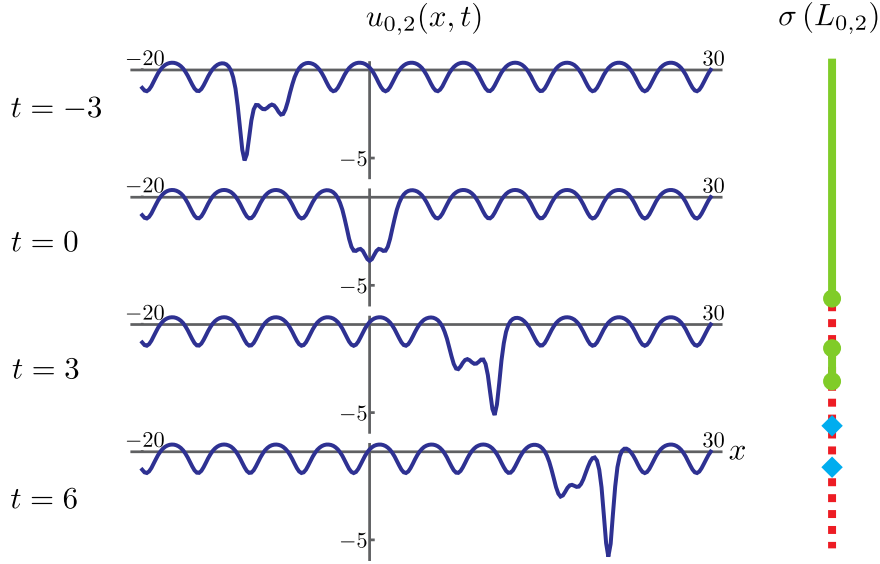}\\
	\caption{A pair of pulse solitons  in the 
	asymptotically periodic background
	of the KdV solution.
	Here $\mu=1$, $k=0.9$, $\beta^-_1=0.8$, 
	$\beta^-_2=0.6$, $C^-_1=C^-_2=1$.
	As in the case of asymptotically free background shown in Fig. 
	\ref{figure1}, a more deep defect 
	propagates with a higher speed.
	Each of the  two defects here
	supports a bound state in the lower 
	forbidden semi-infinite band of the perturbed one-gap 
	Lam\'e system.
	}\label{figure3}
\end{figure}

In  described solutions, an arbitrary number of solitons 
can be eliminated by taking limits of the type 
 $C^-_{j}\rightarrow 0$ or  $C^-_{j}\rightarrow \infty$
 that corresponds to sending
 the $j$-th soliton to minus or plus infinity. 
 This provokes 
 a
  global phase shift $x\rightarrow x-\frac{\mu}{\beta^-_j}$
 for $C^-_{j}\rightarrow 0$,
 or $x\rightarrow x+\frac{\mu}{\beta^-_j}$  
 when $C^-_{j}\rightarrow \infty$,
  in the crystalline  background as well as in the remaining 
  solitons, and additional change in the parameters 
  $C^-_{j'}$, $j'\neq j$, which depend 
on  $\beta^-_{j'}$ and ${\beta^-_j}$, see refs.  \cite{5A,APscat}.

The indicated phase shifts of the crystalline background 
can be understood in the following simple way.
Consider  the first order Darboux transformation based on the 
function $F(x,\beta^-)$  from  Eq. (\ref{Fxtb})
and applied to the stationary periodic KdV solution (\ref{u0Lame}).
Since $\log F(x,\beta^-)=\log \Theta \left(\mu x+\beta^- \left|k\right. \right) -
\log \Theta \left(\mu x \left|k\right. \right)- 
\big(\mu x{\rm z}(\beta^- \left|k\right. )-\pi(\beta^-+i{\rm {\bf K}}'\vert k)  t\big)$,
and $-2\left(\log \Theta \left(\mu x \left|k\right. \right)\right)_{xx}=u_{0,0}(x)$,
we see that the Darboux transformation generated by $F(x,\beta^-)$ 
transforms $u_{0,0}(x)$ into $u_{0,0}(x+\beta^-/\mu)$,
while the transformation based on 
$F(x,-\beta^-)$ produces  the displacement 
$x\rightarrow x-\beta^-/\mu$. When 
in the Darboux-Crum transformation (\ref{u0lxt})
we take the limit  $C^-_j\rightarrow 0$, 
in the function (\ref{Fdefpmj}) the term with 
$F(x,-\beta^-_j)$ survives.  
Due to relation ${\rm z}(-\beta^-_j\vert k)<0$,
this function exponentially increases  in the region 
$x\rightarrow +\infty$, 
and this provokes the displacement 
$x\rightarrow x-\beta^-/\mu$ of the asymptotically
periodic background of the solution in that region. 
Analogously,  for the limit $C^-_j\rightarrow \infty$,
the first  term with $F(x,\beta^-_j)$ survives 
in the function (\ref{Fdefpmj}), and 
the asymptotically periodic background in the solution 
will be displaced by the shift $x\rightarrow x+\beta^-/\mu$ 
in the region to the left ($x\rightarrow -\infty$) 
from  the rest of the  surviving solitons.

\subsection{ Solitons of the compression modulation type}\label{case ii)}

\begin{figure}[h]
	\centering
	\includegraphics[scale=2.7]{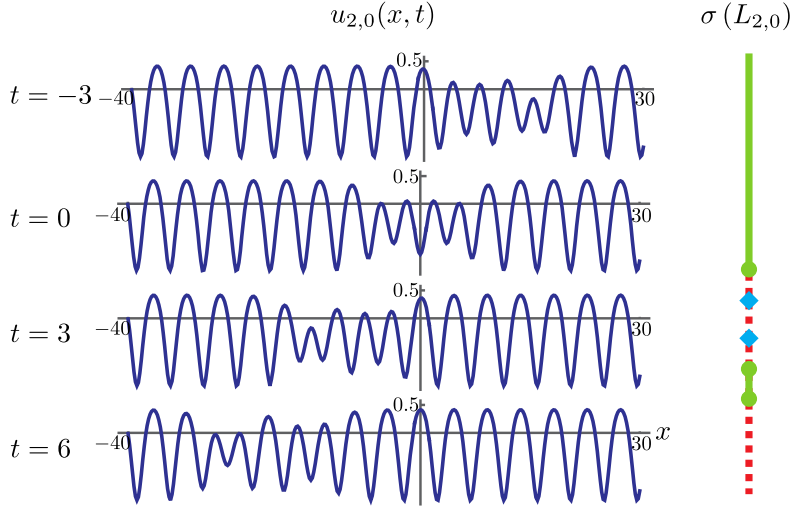}\\
	\caption{A pair of solitons of the compression modulation type propagating to the left 
	 in a stationary crystalline background. 
	 Here
	 $\mu=1$, $k=0.9$, $\beta^+_1=1$, $\beta^+_2=1.3$, $C^+_1=C^+_2=1$.
	 Due to  the  presence of the valence band in the spectrum of Lam\'e system
	 $L_{0,0}$,
	 defects of this  type  have no direct analogs in the asymptotically  
	 free, homogeneous
	 background case.
	 }\label{figure4}
\end{figure}

The states (\ref{Psixt}) with $\alpha=\pm \beta^+$, $0<\beta^+<{\rm {\bf K}}$,
which we denote here as $\Phi(x,t,\pm \beta^+)$,
\be\label{Phibeta+}
	\Phi(x,t,\beta^+)=
	\frac{{\rm H}\left(\mu x+\beta^+ \vert k\right) }{\Theta
	\left(\mu x\vert k\right)}
	\exp\Big(-\mu x\, {\rm Z}(\beta^+ \vert k )+\pi(\beta^+\vert k)t\Big)\,,
\ee
where ${\rm Z}(\beta^+ \vert k )>0$ and 
\be\label{pibeta+def}
	\pi(\beta^+\vert k)=-4 \mu^3 k^2  \,\text{sn}
	\left(\beta^+ \left|k\right.\right) \text{cn}\left(\beta^+ \left|k\right.\right) 
	\text{dn}\left(\beta^+ \left|k\right.\right)<0\,,
\ee
correspond to the energy gap (finite, upper forbidden band) 
$E_1<E(\alpha)<E_2$, see Eq. (\ref{E0E1E2}),
 in the spectrum of 
the  Lam\'e system. They are represented by real functions, and can be used 
in the Darboux transformation (\ref{D1}).
These functions, however, have infinite number of zeroes on the real line, 
and their use would produce a singular transformation.
 To resolve this problem, in the simplest case we can realize 
 a
 second order $(n=2)$
 Darboux-Crum transformation (\ref{uxtn}),
 ${u}_{2,0}(x,t)=u_{0,0}(x)-2(\log W(\Phi_+(1),\Phi_-(2)))_{xx}$,
by taking $\Phi_+(1)=C^+_1\Phi(x,t,\beta^+_1)
	+\frac{1}{C^+_1}\Phi(x,t,-\beta^+_1)$ and
	$\Phi_-(2)=C^+_2\Phi(x,t,\beta^+_l)
	-\frac{1}{C^+_2}\Phi(x,t,-\beta^+_2)$, with 
	$0<\beta^+_1<\beta^+_2<{\rm {\bf K}}$ and $C^+_{1,2}>0$.
The non-singular potential  ${u}_{2,0}(x,t)$ supports two bound 
states in the spectrum of the Schr\"odinger operator $L$
with energy values given by (\ref{lama}), which
are  in the gap, 
between the two continuous bands of the Lam\'e system. 
The solution ${u}_{2,0}(x,t)$ of the KdV equation, 
depicted in Figure \ref{figure4}, 
describes the propagation of the two solitons of the 
compression modulation type in the asymptotically periodic background.
They are similar to  the so-called grey solitons, see 
\cite{Kivshar,GraySol,GrayWater}.
The described Darboux-Crum procedure can be 
generalized to obtain the potential supporting $2l$ bound states in the gap,
\be\label{u2l0}
 	{u}_{2l,0}(x,t)=u_{0,0}(x)-2(\log W(\Phi_+(1),
	\Phi_-(2),...,\Phi_+(2l-1),\Phi_-(2l)))_{xx}\,,
\ee
\be\label{Phipmjdef}
	\Phi_{(-1)^{j+1}}(j)=C^+_j\Phi(x,t,\beta^+_j)
	+(-1)^{j+1} \frac{1}{C^+_j}\Phi(x,t,-\beta^+_j)\,,\qquad
	j=1,\ldots,2l\,,
\ee
\be
	0<\beta^+_1<\beta^+_2<...<\beta^+_l<{\rm {\bf K}},
	\quad 0<C^+_j<\infty\,.
\ee
The solution $	{u}_{2l,0}(x,t)$ has 
$2l$ compression modulation type solitons which move \emph{to the left}
in the stationary crystalline background.

In definition of the  wave functions 
(\ref{Fdefpmj}) and (\ref{Phipmjdef}), which are linear
combinations of the  eigenstates of the Lax operator 
$P$ with opposite eigenvalues $\pi(\alpha)$ and $-\pi(\alpha)$,
the dependence on the evolution parameter 
$t$ can be transferred from the common eigenstates 
of $L$ and $P$ into corresponding coefficients
$C^-_j$ and $C^+_j$. In this way, function (\ref{Phipmjdef})
can be presented equivalently in the form
$\Phi_{(-1)^{j+1}}(j)=C^+_j(t)\Phi(x,t=0,\beta^+_j)
	+(-1)^{j+1} \frac{1}{C^+_j(t)}\Phi(x,t=0,-\beta^+_j)$,
where $C^+_j(t)=C^+_j\exp (\pi(\beta^+_j)t)$, 
and  $\pi(\beta^+_j)$ is 
given by Eq. (\ref{pibeta+def}) with $\beta^+$ changed for 
$\beta^+_j$.
For (\ref{Fdefpmj}) we have analogous 
equivalent representation with $\pi(\beta^-+i{\rm{\bf K}}')$
given by Eq. (\ref{pibeta-}).

To obtain the solution with odd number of solitons of the
compression modulation  type,
there are the following  possibilities.
We can choose one of the arbitrary constants 
$C^+_j$ and apply to the solution (\ref{u2l0})  the limit  $C^+_j\rightarrow 0$
(or  $C^+_j\rightarrow \infty$).
Such a  limit results in sending the $j$-th soliton to $x=-\infty$ (or $x=+\infty$) 
without affecting the rest of solitons except inducing 
a 
global phase shift  in them and in 
the background 
(equal to $x\rightarrow x-\frac{\beta^+_j}{\mu}$ for $C^+_j\rightarrow 0$
and $x\rightarrow x+\frac{\beta^+_j}{\mu}$ for $C^+_j\rightarrow \infty$)
due to 
the
 nonlinear interaction in the KdV equation, 
and additional change in  constants $C^+_{j'}$, $j'\neq j$, 
depending on $\beta^+_{j'}$ and $\beta^+_j$, see  \cite{APscat}.
Another option to generate the solution with odd number of solitons of the
compression modulation type is to take
$\beta^+_1=0$ (in this case, $\Phi_+(1)\propto\text{sn}(\mu x|k)$),  or 
$\beta^+_{2l}={\rm{\bf K}}$ (then,  $\Phi_-(2l)\propto\text{cn}(\mu x|k)$).
One of the indicated  states corresponds to 
the edge of the conduction band of the one-gap Lam\'e system ($\beta^+_1=0$),
while another one corresponds  to the upper edge 
of the valence band ($\beta^+_{2l}={\rm{\bf K}}$).
The same effect can be obtained just
by applying  the limit $\beta^+_1\rightarrow 0$, or
  $\beta^+_{2l}\rightarrow {\rm{\bf K}}$ in (\ref{u2l0}), that 
  results in eliminating  the corresponding bound state
  from the spectrum of  
  the Schr\"odinger system 
  $L=-\frac{d^2}{dx^2}+u_{2l,0}(x,t)$, see ref. \cite{5A}.
 The issue of velocities for the described solutions 
 will be discussed below, in Section   \ref{veloSec}.

\subsection{ Mixed case}\label{case iii)}

\begin{figure}[h]
	\centering
	\includegraphics[scale=3]{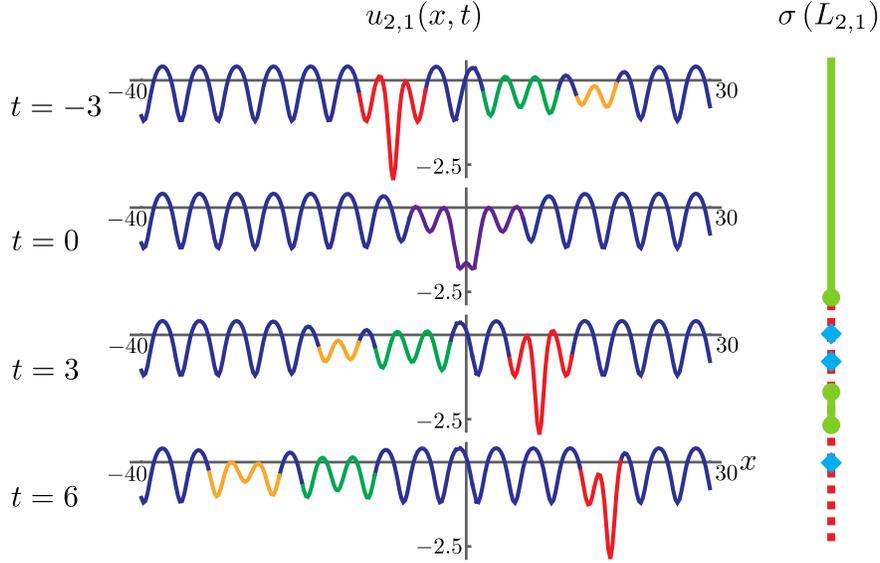}\\
	\caption{The KdV solution with one pulse soliton 
	(given by $\beta^-_1$ and shown in  red) 
	propagating to the right,
	and two solitons of the compression modulation type 
	(characterized by $\beta^+_1$ and $\beta^+_2$ and depicted in 
	 orange and  green, respectively)
	moving to the left in a stationary crystalline background.
	 More rapid compression soliton ($\beta^+_1$)
	 supports a  bound state of higher energy
	 in the spectrum of the  associated Schr\"odinger system  $L_{2,1}=-\frac{d^2}{dx^2}+u_{2,1}(x,t)$. 
	At the moment $t=0$, solitons are in the zone of a strong interaction, shown in violet,
	and are not well separated. 
	The
	colour highlighting used here  is rather conditional 
	since the defects  are transformed into a 
	background asymptotically.  The parameters are chosen so that 
	the solution at $t=0$ is symmetric with respect to the point $x=0$.
	 Here, $\mu=1$, $k=0.9$, $\beta^-_1=0.9$, $\beta^+_1=1$, 
	 $\beta^+_2=1.3$,
	  $C^-_1=C^+_{1,2}=1$.   }\label{figure5}
\end{figure}

In the mixed  case, the Darboux-Crum transformation takes the form
\be\label{u2lm}
 	{u}_{2l,m}(x,t)=u_{0,0}(x)-2(\log W(\Phi_+(1),\Phi_-(2),...,\Phi_-(2l),\mathcal{F}_+(1),
	 \mathcal{F}_-(2),...,\mathcal{F}_{(-1)^{m}}(j)))_{xx}\,.
\ee
This solution has 
$2l$ solitons of the compression modulation type,
which move to the left, and  $m$ solitons of the potential well type, 
which propagate to the right.
In the associated Schr\"odinger system 
$L_{2l,m}=-\frac{d^2}{dx^2}+u_{2l,m}(x,t)$, the potential well 
and compression modulation type
solitons support the bound states in the lower forbidden band and in the gap  
of the energy spectrum, respectively. In Figure \ref{figure5}, one can see how 
solitons of different types propagate in opposite directions over the 
stationary background. 

Here, the case with the odd number of the solitons of the 
compression modulation type can be obtained from (\ref{u2lm})
in the way described in the preceding  Section~\ref{case ii)}.

\subsection{On  velocities and amplitudes of solitons in crystalline 
background}\label{veloSec}

For asymptotically free solutions discussed in 
Section \ref{multisolKdV}, 
the
multi-soliton solution can be presented in the form 
(\ref{sol_n}), which is similar to the form of the second term  
on the right in (\ref{u0lxt}), (\ref{u2l0}), or  (\ref{u2lm}).
Hyperbolic $\cosh$ and $\sinh$ functions in (\ref{sol_n}) 
are just the linear combinations of 
exponents 
$\exp(\kappa_j(x-4\kappa_j^2 t))$
and $\exp(-\kappa_j(x-4\kappa_j^2 t))$ with constant coefficients.
As a result, when solitons are well separated
(this happens for sufficiently large values of $\vert t\vert $),
the solution  (\ref{sol_n}) reduces 
to the sum of one-soliton solutions, $u_n(x,t)\approx \sum_{j=1}^n -
2\kappa_j^2\sech^2\left(\kappa_j(x-\tilde{x}_{0j}-4\kappa_j^2 t)\right)$,
with obviously  identified velocities and amplitudes. 
Since the exponents in functions (\ref{Fxtb}) 
and (\ref{Phibeta+}) are odd functions 
of the parameters $\beta^-$ and $\beta^+$, 
respectively, functions (\ref{Fdefpmj})
and (\ref{Phipmjdef}) similarly 
to the hyperbolic functions are linear combinations
of the exponents with arguments of the form 
$\varphi(x-\cV t)$ and $-\varphi(x-\cV t)$, but  now with the 
coefficients which are  certain periodic functions in~$x$.
As a consequence, 
unlike the case of 
KdV solutions over the asymptotically 
free background, it is not so obvious how to define 
amplitudes and velocities of solitons propagating in the 
crystalline background. 
Even when defects are well separated,
their velocities and amplitudes  are varied 
at each instant  of time due to their nonlinear, position-dependent 
interaction with the oscillating background.
Nevertheless, by analogy with the asymptotically free case, 
 one can observe that 
the quantities $\mu^2{\rm z}^2(\beta^-\vert k)$ and 
$\mu^2{\rm Z}^2(\beta^+\vert k)$ give us 
a relevant information 
on amplitudes of the well separated pulse
and modulation types defects, respectively, provided that
for a  well separated soliton defect the 
corresponding pre-exponential periodic  factor in (\ref{Fxtb}) 
or  (\ref{Phibeta+}),
associated with the  crystalline background, will
be less significant than the 
exponential factors there.
This observation means a monotonic increase
of the amplitude of the pulse type soliton 
with increasing energy modulus of the trapped 
by it bound state in the spectrum of the associated perturbed 
one-gap Lam\'e system.
Such a monotonic increase will be valid, however,  starting 
from some sufficiently low energy value
inside the lower forbidden band of the spectrum. 
For soliton defects of the compression 
modulation type 
the
 picture is more complicated
since  in the interval $0 < \beta^+ < {\rm {\bf K}}$ 
the function ${\rm Z}^2(\beta^+\vert k)$ takes zero value  
when $\beta^+$ tends to zero, then monotonically increases  till some 
maximum value at some $\beta^+_*$ inside the indicated interval,
and then monotonically decreases taking 
zero  value 
at the edge  $\beta^+={\rm {\bf K}}$.  
So, we can only conclude that the amplitude of the compression
 modulation type defect will tend to zero when the energy of the 
 bound state trapped by it approximates the edges of the gap in
  the spectrum of the perturbed one-gap
 Lam\'e system,
and will take some maximal value at some energy
of the corresponding  trapped bound state inside 
the gap. 
With the same reservations
one  can say that  the width of the 
defects is proportional to $1/\mu{\rm z}(\beta^-)$
and $1/\mu{\rm Z}(\beta^+)$ for the pulse and 
modulation type solitons, respectively.

To define the velocities of the solitons,
we consider the case of a well separated defect
of the pulse type, while for the 
 compression modulation 
type defect the reasoning will be similar.

The  well separated pulse soliton 
with index $j$ 
will have the form described by the one-soliton solution 
$u_{0,1}$  given by Eq. (\ref{u0lxt}) 
with some  parameter $C^-_1$ and $\beta^-_1=\beta^-_j$, i.e. 
this soliton will depend only on  eigenstate
 $\mathcal{F}_+(j)$ with certain parameter $C^-_j$.
Then we can identify the velocity of this soliton from the 
condition $u_{0,1}(x,t)=u_{0,1}(x+\Delta_x,t+\Delta_t)$
to be valid for all $x$ and $t$ for which soliton 
remains  well separated from other defects,
where $\Delta_x$ and $\Delta_t$ are some constants.
This condition has a solution with 
$\Delta_x=nT$, where $T=\frac{2{\rm {\bf K}}}{\mu}$ 
corresponds to  the period of the 
asymptotically periodic crystalline background 
and $\Delta_t=n{2{\rm {\bf K}}{\rm z}(\beta^-_j\vert k)}/({\mu \pi(\beta^-_j+
i{\rm {\bf K}}'\vert k)})$, while $n$ is any  integer
not violating the condition of separation of  the
soliton.
{}From here we identify the velocity of the pulse soliton 
as  $\cV(\beta^-_j)=\Delta_x/\Delta_t$,
and obtain 
 \be\label{Vpulse}
	\mathcal{V}(\beta^-_j)=\frac{4 \mu^2 \frac{ \text{cn}\left(\beta^-_j\left|k\right.\right) 
		\text{dn}\left(\beta^-_j\left|k\right.\right)}{\text{sn}^3\left(\beta^-_j\left|k\right.\right)} }
	{
	{\rm Z}\big(\beta^-_j \left|k\right. \big)+
	\frac{\text{cn}\left(\beta^-_j\left|k\right.\right) \text{dn}\left(\beta^-_j\left|k\right.\right)}
	{\text{sn}\left(\beta^-_j\left|k\right.\right)}
	}>0\,.
\ee
This corresponds  to the velocity
with which a fixed value of the argument 
of the exponent in (\ref{Fxtb}) propagates in space and time,
$\varphi(x-\cV t)=const$.

In a similar way, for the compression modulation type defect we find
\be\label{Vbeta+j}
	\mathcal{V}(\beta^+_j)=-4 \mu^2 k^2  \frac{\text{sn}
	(\beta^+_j \left|k\right.) \text{cn}(\beta^+_j \left|k\right.) 
	\text{dn}(\beta^+_j \left|k\right.)}{{\rm Z}(\beta^+_j \vert k 
	)}<0\,.
\ee

For $\beta^-_j\rightarrow 0$, the energy of the bound state trapped by the 
pulse defect tends to minus infinity, and the amplitude and velocity
(\ref{Vpulse}) of the soliton tend to infinity, 
while its width tends to zero.
For $\beta^-_j\rightarrow {\rm {\bf K}}$,
the amplitude of the soliton tends to zero, 
its width tends to infinity, while the limit 
of (\ref{Vpulse}) is finite and equals
\be\label{Vbeta-lim}
	\lim_{\beta^-_j\rightarrow {\rm {\bf K}}} 
	\cV (\beta^-_j)=4\mu^2k'^2\frac{{\rm {\bf K}}}{{\rm {\bf E}}}\,.
\ee
Here ${\rm {\bf E}}$ is the complete elliptic integral of the second kind,
and as a consequence of the inequality relations 
$\frac{1}{k'^2}>\frac{{\rm {\bf K}}}{{\rm {\bf E}}}>1$ \cite{PAN}, 
one finds that the limit value (\ref{Vbeta-lim}) 
is inside the interval $(4\mu^2k'^2,4\mu^2)$.

Similarly, for the compression modulation type defect,
for the limits when its amplitude tends to zero, its velocity
tends  to nonzero limits,
\be
 	\lim\limits_{\beta^+_j\rightarrow0} \mathcal{V}(\beta^+_j)=-4  
	 \mu^2k^2\frac{ {\rm {\bf K}}}{{\rm {\bf K}}-{\rm {\bf E}}}<0\,,
	 \qquad
	 \lim\limits_{\beta^+_j\rightarrow {\rm {\bf K}}} \mathcal{V}
	 (\beta^+_j)=-4\mu ^2 k^2 k'^2 \frac{ {\rm {\bf K}} }{{\rm {\bf E}}-k'^2{\rm {\bf K}} }<0\,,
\ee
while its width tends to infinity. 
 
\subsection{Galilean symmetry}\label{GalSec}

The described solutions  can be modified by employing the Galilean symmetry 
of the KdV equation.
This symmetry means that if 
 $u(x,t)$ is a solution of the KdV equation $u_t=6uu_x-u_{xxx}$, 
 then  ${u}_G(x,t)=u(x-6Gt,t)-G$ is also a solution of the KdV equation
 for any choice of a real constant $G$. 
 In the described solutions,
 the crystalline background over which the soliton defects propagate deforming it
due to a non-linear 
 interaction,
 was static. The application  of Galilean transformations
 to the solutions will boost the defects and background and also shift 
 vertically the solutions 
 by the additive constant $-G$.
 This Galilean transformation 
 will not change, however, the relative velocities between defects 
 and their velocities with respect to the boosted crystalline background. 
As we shall see below, the Galilean symmetry of the KdV equation 
will play a crucial 
role in the construction of the solutions for the mKdV equation
by means of the Miura-Darboux-Crum transformations.

\section{Unification of the
KdV and mKdV equations by Miura-Darboux-Crum  
transformations, and exotic supersymmetry}\label{KdVmKdVuni}

\subsection{Miura-Darboux-Crum  transformations}\label{MiuDarCr}
The defocusing 
mKdV equation and the KdV equation,
\be\label{mKdV_KdV}
	v_t-6v^2v_x+v_{xxx}=0\,,
	\qquad u_t-6uu_x+u_{xxx}=0\,,
\ee
are related by the Miura transformation 
\cite{Miura}. Namely, 
the substitution of
$
	u^\pm=v^2\pm v_x
$
into the KdV equation gives
\be\label{u+u-}
	u^\pm_t-6u^\pm u^\pm_x+u^\pm_{xxx}=
	(2v\pm\partial_x)(v_t-6v^2v_x+v_{xxx})\,.
\ee
The two relations in (\ref{u+u-})  mean that  if $v$ is a solution of the mKdV equation,
then both $u^\pm$ satisfy the KdV equation. On the other hand,
if both $u^+$ and  $u^-$  are solutions of the KdV equation,
and there exists $v$ such that $u^\pm=v^2\pm v_x$, 
then $v$ obeys the mKdV equation \cite{AGPfer}.
Unlike the KdV case, if 
$v(x,t)$ is the mKdV solution, then  $-v(x,t)$ is also solution.

{}From the chains of the Darboux-Crum transformations 
we know   that 
the function $V_m(x,t)$ in (\ref{vdef})
 and the 
corresponding solutions of the KdV equation 
${u}_m(x,t)$ and $u_{m-1}(x,t)$ 
are related by 
\be\label{vmumum-1}
	V_m^2(x,t)+(V_{m}(x,t))_x=u_{m-1}(x,t)-\lambda_m\,,
	\quad V_m^2(x,t)-(V_{m}(x,t))_x={u}_m(x,t)-\lambda_m\,.
\ee
With taking into account the Galilean symmetry of the KdV equation,
we displace   $x\rightarrow x-6\lambda_m t$,   and 
make a change 
 $u(x,t)\rightarrow
{u}(x-6\lambda t,t)-\lambda\equiv {u}^\lambda(x,t)$. 
Eq. (\ref{vmumum-1}) will transform  then into 
\be\label{DCvmum}
 	v_m^2(x,t)+ (v_{m}(x,t))_x={u}^{\lambda_m}_{m-1}(x,t)\,,
	\qquad  v_m^2(x,t)
 	- (v_{m}(x,t))_x={u}^{\lambda_m}_m(x,t)\,,
\ee
where  $v_m(x,t)=V_m(x-6\lambda_m t,t)$.
The  $v_m(x,t)$ are therefore 
solutions of the mKdV equation. 

Utilizing this observation and 
solutions of the KdV equation obtained by means of 
the Darboux-Crum transformations, we can construct 
infinite number of solutions for the mKdV equation.
The KdV solutions from Section \ref{multisolKdV} will provide us with
the mKdV solitons  taking constant asymptotic values at infinity.
On the basis of the KdV multi-soliton defects from the preceding Section,
we will obtain solutions for the mKdV equation in the form of 
soliton defects propagating in the crystalline background.

\subsection{Exotic  $N=4$ nonlinear supersymmetry}\label{SexoticSUSY}
 
Before we proceed to the discussion of the solutions to the mKdV equation
and their peculiarities, 
 we note that  on the basis of relations (\ref{DCvmum}), 
 the corresponding  solutions of the KdV and mKdV equations can be related 
 by exotic $N=4$ nonlinear supersymmetry.  Besides an ordinary $N=2$ supersymmetry,
 it incorporates two non-trivial, Lax-Novikov integrals for the 
 two Schr\"odinger subsystems associated 
 with the KdV equation, as well as two additional supercharges
 to be higher order matrix differential operators.

 To reveal the exotic supersymmetric structure,
 we first note that an ordinary  $N=2$ supersymmetry in the form of superalgebra 
 \be\label{susyN=2}
 [ \mathcal{L}_m,\mathcal{S}_{m,a}]=0\,, \qquad
 \{\mathcal{S}_{m,a},
 \mathcal{S}_{m,b}\}=2 \mathcal{L}_m\delta_{ab}\,,
 \qquad a,b=1,2\,,
 \ee
 with $\sigma_3$ identified as a $\Z_2$ grading operator,
 is generated 
 by the extended, matrix Schr\"odinger Hamiltonian 
 $\mathcal{L}_m$ and 
 the associated
  Dirac Hamiltonian 	$\mathcal{D}_{m}$,
\be\label{LSLD}
	  \mathcal{L}_m=\left(\begin{array}{cc}
	  L_{m-1}(m)& 0\\ 0& 
 	 L_{m}(m)\end{array}\right)\,,
	 \qquad
 	\mathcal{D}_{m}=\left(\begin{array}{cc}0& A^\dagger_m(m)\\
 	A_m(m)& 0\end{array}\right)\,,
 \ee
 where $\mathcal{S}_{m,1}=\mathcal{D}_{m}$
 and  $\mathcal{S}_{m,2}=
 i\sigma_3 \mathcal{D}_{m}$.
 The Schr\"odinger operators 
 $L_{m-1}(m)$ and  $L_m(m)$,
  $L_n(m)=-\frac{d^2}{dx^2}+
 u_n^{\lambda_m}(x,t)$, $n=m-1,m$,
 are related here by the 
 Darboux (Miura) transformation 
 generated by the first order operators $A_m(m)$
 and $A_m^\dagger(m)$,
 where $A_m(n)=\frac{d}{dx}-V_m(x-6\lambda_nt,t)$,
 and the solution of the mKdV equation 
 $v_m(x,t)=V_m(x-6\lambda_m t,t)$
 can be considered  
 as a superpotential. 
 
 Let us remind  that for
 the quantum mechanical operators $ \mathcal{L}_m$ and 
 $ \mathcal{D}_m$,
parameter $t$ corresponds not  to the Schr\"odinger or 
Dirac 
evolution, but, instead, it is associated here with the
coherent  peculiar isospectral deformations of their  
 potentials  governed  by the 
 KdV and mKdV equations. 
 Any Schr\"odinger system with 
 multi-soliton or finite-gap potential can be characterized by 
 a nontrivial integral of motion in the form of 
 the Lax-Novikov higher derivative  differential 
 operator of odd order \cite{NMPZ,AlgGeo,BurChau}. 
As a consequence,  for the extended,  matrix  system $ \mathcal{L}_m$
 composed from a pair of such peculiar
 Schr\"odinger subsystems related by 
 a Darboux transform,    the  
 quantum mechanical $N=2$  supersymmetry
 associated with  fermionic integrals  $\mathcal{S}_{m,a}$
 is extended up to the exotic $N=4$ nonlinear 
 supersymmetry incorporating two additional
 supercharges $\cQ_{m,a}$.  Additional supercharges 
 are composed from higher (even) order differential operators 
 which intertwine the diagonal elements 
 in $ \mathcal{L}_m$,   and together with building blocks 
 of supercharges $\mathcal{S}_{m,a}$ (being 
 the first order differential operators $A_m(m)$ and 
 $A^\dagger_m(m)$)  factorize effectively the 
 Lax-Novikov integrals of the extended 
 Schr\"odinger system $ \mathcal{L}_m$.
 For the discussion of a general structure of the exotic 
 $N=4$ nonlinear 
 supersymmetry associated with finite-gap and soliton systems see refs. 
  \cite{CJNP,PAN,AGPfer,APscat,CJP2,AGP1} and further references therein. 
 
Specifically, here the extension of $N=2$ supersymmetry
up to the exotic $N=4$ nonlinear supersymmetry
happens as follows. 
 We have started with the 
stationary solution (\ref{u0Lame})   for the KdV equation 
to construct solutions $u_m(x,t)$.
For the initial  Schr\"odinger operator $L_0=-\frac{d^2}{dx^2}+
u_0(x)$, $u_0(x)=u_{0,0}(x)$, we can construct the first order operator 
(\ref{A1def}), $A_1(x,t)=\frac{d}{dx}-v_1(x,t)$,
which provides us with the intertwining relation
$A_1L_0=L_1 A_1$, where $L_1=-\frac{d^2}{dx^2}+
u_1(x,t)$. The Lax operator $P_0=P(u_0)$ constructed on the basis 
of the stationary solution $u_0(x)$ satisfies 
equation (\ref{LP0}),  $[P_0,L_0]=0$. Being the third order
differential operator, 
this is the Lax-Novikov integral 
for the Schr\"odinger system $L_0$. 
The intertwining relation $A_1L_0=L_1 A_1$ 
and the conjugate relation show  that for the 
Schr\"odinger system $L_1$, 
the fifth order differential operator
$P_1(x,t)=A_1P_0A_1^\dagger$
is the Lax-Novikov integral of motion, 
$[P_1,L_1]=0$. Applying  
then Galilean transformation 
 with the parameter $G=\lambda_1$
 to the KdV solutions 
 $u_0(x)$ and $u_1(x,t)$, 
 we obtain the fifth order operators 
 $P_1(1)=A_1(1)P_0(x-6\lambda_1 t)A^\dagger_1(1)$
 and $P_0(1)=L_0(1)P_0(x-6\lambda_1 t)=
 A_1^\dagger(1)A_1(1)P_0(x-6\lambda_1 t)$,
 which are integrals for $L_1(1)$ and $L_0(1)$,
 respectively. Being the product 
 of the integral $P_0(x-6\lambda_1 t)$ with
 the Schr\"odinger Hamiltonian $L_0(1)$,
 in this case the integral 
 $P_0(1)$ is reducible.
 We take it, however,  to construct 
 two bosonic  integrals for the extended system $\mathcal{L}_1$,
\be\label{P1a}
	\mathcal{P}_{1,1}=\left(\begin{array}{cc}P_{0}(1)& 0\\ 0& 
	P_{1}(1)\end{array}\right),\quad 
	\mathcal{P}_{1,2}=\sigma_3\cP_{1,1}\,,
\ee
with the matrix elements to be differential operators 
of the same order.

To get the analogs of integrals (\ref{P1a}) for a general case 
corresponding to the extended Schr\"odinger system 
described by $\cL_m$, we have to change 
the fifth order differential operator $A_1P_0A_1^\dagger$ for 
$\A_mP_0\A_m^\dagger$
that is a differential operator of order $2m+3$ commuting with 
the Schr\"odinger operator $L_m=-\frac{d^2}{dx^2}+u_m(x,t)$.
Then  we realize  a usual Galilean transformation with the velocity
 $-6\lambda_m$ to  obtain the operator
$P_{m}(m)=\A_m(m)P_0(x-6\lambda_m t)\A_m^\dagger(m)$, 
that  is the  Lax-Novikov integral 
for the quantum system $L_{m}(m)$.
In a similar way, as analog of 
 $L_{0}(1)P_0(x-6\lambda_1 t)$ we get
operator $P_{m-1}(m)=L_{m-1}(m
 )\A_{m-1}(m)P_0(x-6\lambda_m t)\A_{m-1}^\dagger(m)$
 that commutes with $L_{m-1}(m)$, where 
$\A_n(m)=A_n(m)A_{n-1}(m)...A_1(m)$. 
In this manner we obtain a pair of matrix operators 
$\cP_{m,1}={\rm diag}\,(P_{m-1}(m),P_{m}(m))$ 
and $\cP_{m,2}=\sigma_3\cP_{m,1}$, which are the integrals 
for the extended Schr\"odinger system 
$\cL_{m}$, and in addition to 
(\ref{susyN=2}), 
we get the relations 
\be
[\cP_{m,a},\cL_{m}]=
[\cP_{m,a},\cP_{m,b}]=[\cP_{m,1},\cS_{m,b}]=0\,.
\ee
At the same time, the commutator of 
$\cP_{m,2}$  with  $\cS_{m,a}$, $a=1,2$, 
will supply us with a pair of new fermionic integrals 
$\cQ_{m,a}$, which are differential operators of even order  
$2m$, that  also commute with 
$\cP_{m,1}$, $[\cP_{m,1},\cQ_{m,b}]=0$.
So, the operator  $\cP_{m,1}$, like $\cL_m$,  is the bosonic central charge
of the superalgebra.  
Together, two bosonic integrals  $\cP_{m,a}$, $a=1,2$,  
allow us to distinguish the eigenstates 
corresponding to the four-fold degenerate eigenvalues
inside the continuous allowed bands of $\cL_m$. 
Also, they detect all the edge-states and all the bound states  
in the spectrum of $\cL_m$ by annihilating them. 
The fermionic integrals  
$\cS_{m,a}$ and  $\cQ_{m,a}$ 
generate transformations between the `up' and `down'
eigenstates of the same eigenvalues 
in the spectrum of $\cL_m$,
and as usually, complex  linear combinations 
of $\cS_{m,1}$ and $\cS_{m,2}$, and of 
$\cQ_{m,1}$
and $\cQ_{m,2}$ will be creation and annihilation 
type 
operators for those eigenstates.
The
bosonic integral $\cP_{m,2}$ 
generates a kind of rotation between supercharges 
$\cS_a$ and $\cQ_a$ \cite{5A,AGPfer,APscat}.

The four fermionic supercharges $\cS_{m,a}$ and 
$\cQ_{m,a}$
and two bosonic integrals $\cP_{m,a}$ together with 
the matrix Schr\"odinger Hamiltonian $\cL_m$
generate 
the
exotic nonlinear $N=4$ supersymmetry,
whose superalgebraic relations will
contain the  coefficients to be polynomials 
in the central charge $\cL_m$. Such unusual nonlinear extension 
of supersymmetric structure related to integrable
systems was discussed in different aspects in 
refs.~\cite{5A,CJNP,PAN,APscat,CJP2,AGP1}. 
The anti-commutation relations 
for $\cS_{m,a}$ in (\ref{susyN=2}), and
similar relations for $\cQ_{m,a}$ with 
right hand side
to be a certain polynomial of order $m$ in $\cL_m$,
together reflect the fact that the square 
of the Lax-Novikov integrals $\cP_{m,a}$,  in correspondence 
with the Burchnall-Chaundy
theorem \cite{BurChau}, is a certain polynomial of odd order 
$(2m+1)$ in $\cL_m$.

There also exist the cases of the systems, the explicit examples of which will be 
considered below, when the described structure of exotic supersymmetry 
can be reduced in the order of differential operators
corresponding to the  set of integrals $\cP_a$ and $\cQ_a$.
This happens when the Schr\"odinger potentials 
$u_m^{\lambda_m}(x,t)$ and  $u_{m-1}^{\lambda_m}(x,t)$ 
are completely isospectral, 
and the ordinary
$N=2$ supersymmetry 
generated by the first order supercharges
$\cS_a$ in accordance with
(\ref{susyN=2})
is spontaneously  broken. 
Specifically, in the case  when 
Schr\"odinger potentials $u_m^{\lambda_m}(x,t)$ and $u_{m-1}^{\lambda_m}(x,t)$ 
have a difference in one bound state in the spectra of the systems 
$L_m(m)$ and $L_{m-1}(m)$, 
the spectrum of the corresponding Dirac 
Hamiltonian with the scalar potential 
$v_m(x,t)$ will contain one kink as a defect.
If this is the case,  there is no reduction 
in the structure of the exotic supersymmetry generators;
the envelope 
of the corresponding
oscillating  eigenfunction 
$\Psi_m(x)$ of the initial 
one-gap Lam\'e system used in the Darboux-Crum  
construction will 
exponentially increase  in both positive and negative directions
of $x$. In contrast, when  the envelope of $\Psi_m(x)$ 
increases exponentially in one direction while 
in other direction  exponentially decreases, 
the corresponding  
first order intertwining operator $A_{m}(m)=X_{m-1}(m)=
\frac{d}{dx}-v_{m-1}(x,t,\lambda_m)$
generates a nonlinear shift in the already present 
soliton defects as well as in the background,
without  adding a bound state into
the spectrum of  $L_{m}(m)\equiv \tilde{L}_{m-1}(m)$ in comparison with that 
of $L_{m-1}(m)$, 
$X_{m-1}(m)L_{m-1}(m)=\tilde{L}_{m-1}(m)X_{m-1}(m)$.
The superpotential 
$v_{m-1}(x,t,\lambda_m)$
relating such a pair of isospectral Schr\"odinger systems 
can always be obtained from the appropriate 
superpotential $v_m^{(as)}(x,t)$ corresponding to the 
associated irreducible extended Schr\"odinger system
by one  of the limits of the form  
$\lim\limits_{C_m\rightarrow 0,\infty} 
v_m^{(as)}(x,t)=\pm v_{m-1}(x,t,\lambda_m)$
\cite{5A,APscat}. The spectrum of the corresponding Dirac Hamiltonian
$\cD_m$ (supercharge $\cS_{m,1}$),
with the scalar potential 
$v_{m-1}(x,t,\lambda_m)$ is symmetric
with  a central gap between bound states or continuous bands,
and it can contain defects of the kink-antikink type only,
but never kink type defects. 
For such extended Schr\"odinger systems,
the differential order of integrals $\cP_a$ and $\cQ_a$ 
will reduce by two.
{}From the point of view of the limit 
of the associated appropriate  irreducible system 
$\cL_{m}^{(as)}$,
we have $\cL_{m}^{(as)}\rightarrow \cL_{m}$,
and then one can find  that 
$\cP_{m}^{(as)}\rightarrow \cL_{m}\cP_{m}$,
and a similar relation for supercharges $\cQ_a$.
The indicated reducibility of the matrix differential operators 
$\cP_a$ and $\cQ_a$ 
does not affect the nature of the 
supersymmetry: the corresponding extended Schr\"odinger
system presented  by the $2\times 2$ diagonal matrix Hamiltonian operator 
$  \mathcal{L}_m$ is characterized by the exotic 
nonlinear $N=4$ supersymmetry generated by  
 two first order supercharges $\cS_a$ and 
by two fermionic integrals $\cQ_a$ being higher (even) order 
differential operators, and by two bosonic integrals 
composed from the Lax-Novikov integrals of the 
completely isospectral Schr\"odinger subsystems.

\section{Soliton solutions for the mKdV equation}\label{mKdVsolSec}

In the next two subsections, we discuss briefly  the mKdV solutions 
corresponding to the multi-kink-antikink solitons propagating over the
asymptotically  free 
kink  or kink-antikink backgrounds, 
and then we consider  much more rich 
case corresponding to solutions in the  crystalline kink and kink-antikink 
backgrounds.

\subsection{Multi-kink-antikink solutions over a kink background}\label{mKdVsub1}

\begin{figure}[h]
	\centering
	\includegraphics[scale=3]{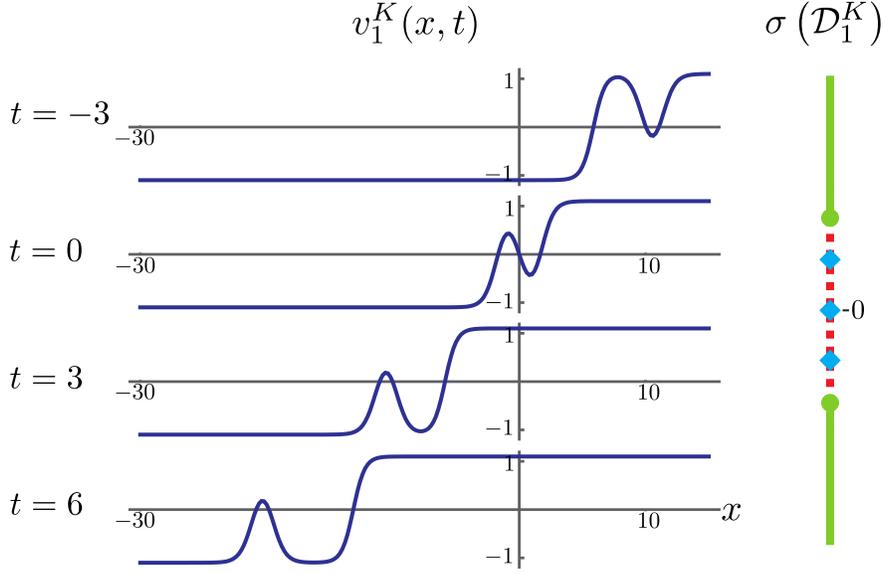}\\
	\caption{Solution of the mKdV equation 
	representing the kink-antikink  (characterized by the 
	$\kappa_1$ parameter)
     propagation over the
	kink (given by $\kappa_2$) background, both moving to the left.
	When the soliton, which is faster,  is to the right 
	from the kink center (see the $t=-3$ instant), it has a form  of the 
	antikink-kink perturbation. Overtaking the kink center,  
	 it flips and transforms into the kink-antikink  
	perturbation 	($t=3$ and $t=6$).
	Here $\kappa_1=1$, $\kappa_2=1.1$, $x_{0i}=0$, $i=1,2$.
	The change  $v\rightarrow -v$ provides  
	a solution over the antikink background with 
	a mutual transform of the antikink-kink and kink-antikink
	perturbations. The kink has a bigger amplitude $2\kappa_2$
	equal to the distance between asymptotes, and this
	value corresponds also to the size of the central gap in the spectrum 
	of  the associated Dirac Hamiltonian operator 
	shown on the right. Filled circles correspond to  non-degenerate energy levels 
	at the edges of the doubly degenerate continuous parts of  the spectrum, 
	while blue squares  indicate 
	non-degenerate bound states inside the spectral gap.
	Zero energy bound state  is associated with the kink,
	the two other bounds states are associated with the kink-antikink
	perturbation in the mKdV solution.}\label{figure6}
\end{figure}

The employment of the  multi-soliton solutions of the KdV equation  constructed
on the basis of  the initial trivial solution $u=0$ allows  
to  find topologically nontrivial solutions $v^K(x,t)$ for the mKdV equation
with the asymptotic behaviour 
$v^K(+\infty,t)=-v^K(-\infty,t)=const\neq 0$.
Index $K$ reflects here the kink-type nature of the 
solutions.
To obtain the mKdV solutions with the indicated asymptotic behaviour, 
we should  take $u_m$ with one more bound state
in the spectrum of the associated Shr\"odinger system in comparison with the
$(m-1)$ bound  states supported by the  
potential $u_{m-1}$. In this case, the solution to the mKdV equation will have 
a form of the multi-kink-antikink defect propagating over the kink. 
The kink-antikink perturbations 
will always  have amplitudes smaller  than the kink amplitude,
characterized by the parameter $\kappa_m$, due to 
the ordering   $0<\kappa_1<\ldots <\kappa_m$. 
The kink-antikinks ($j=1,\ldots,m-1$) and kink ($j=m$)
propagate to the left with 
the velocities $\cV_j=4\kappa_j^2-6\kappa_m^2$. 
So, the kink's speed is the lowest, being equal to
 $2\kappa_m^2$, while the kink-antikink solitons with lower amplitudes 
 have higher speeds. 
Such solutions generalize the mKdV kink solution 
$v^K_0(x,t)=\kappa \tanh \kappa(x+2\kappa^2 t)$, and 
analytically they are given by 
\be\label{vmVm6k}
	v^K_{m-1}(x,t)=V^K_{m}(x +
	6\kappa^2_{m}t,t)\,,
\ee
where,  in correspondence with (\ref{sol_n}),
\be\label{VmOmOm}
	V^K_{m}(x,t)= \Omega^K_{m-1}(x,t)-\Omega^K_m(x,t)\,,\qquad
	\Omega^K_m=
	-\left(\log W(\cosh X^-_1,
	\sinh X^-_2,\ldots, f(X^-_{m}))\right)_x\,,
\ee
$f(X^-_{m})=\sinh X^-_m$  if $m$ is even,
$f(X^-_m)=\cosh X^-_m$ if $m$ is odd,
and $X^-_m$  is defined in (\ref{sol_n}).
Functions   $\pm v^K_{m}(x,t)$, 
representing solutions for
the defocussing 
mKdV equation, are  of the kink type with $m$ solitons that deform 
in their propagation the moving kink  (or antikink)   background 
without overpassing its asymptotes. 
An example of such solution is represented  in Fig.~\ref{figure6}.

\subsection{Multi-kink-antikink solutions over the topologically trivial 
background}\label{mKdVsub2}

 The multi-kink-antikink solutions propagating over the topologically trivial background 
 with asymptotic behaviour $v(-\infty, t)=v(+\infty,t)=const \neq 0$  
 can be obtained from the KdV solutions 
$u_m$ and  $u_{m-1}$ 
which, as the Schr\"odinger potentials, support 
 the same number [($m-1$), $m=2,\ldots$]
 of the bound states 
with coinciding energies. 
In such a  pair,  potentials $u_m$ and  $u_{m-1}$ are related by the  
Darboux transformation that displaces  solitons \cite{5A,APscat}.
They generalize the simplest kink-antikink 
solution  for the mKdV equation,
\be\label{v0xt}
	v_0(x,t)=-\kappa_1\,,
\ee
which corresponds to the nonzero mass term $m=\kappa_1$
of the free Dirac Hamiltonian operator $\mathcal{D}_{0}$,
see  (\ref{LSLD}).

\begin{figure}[h]
	\centering
	\includegraphics[scale=3]{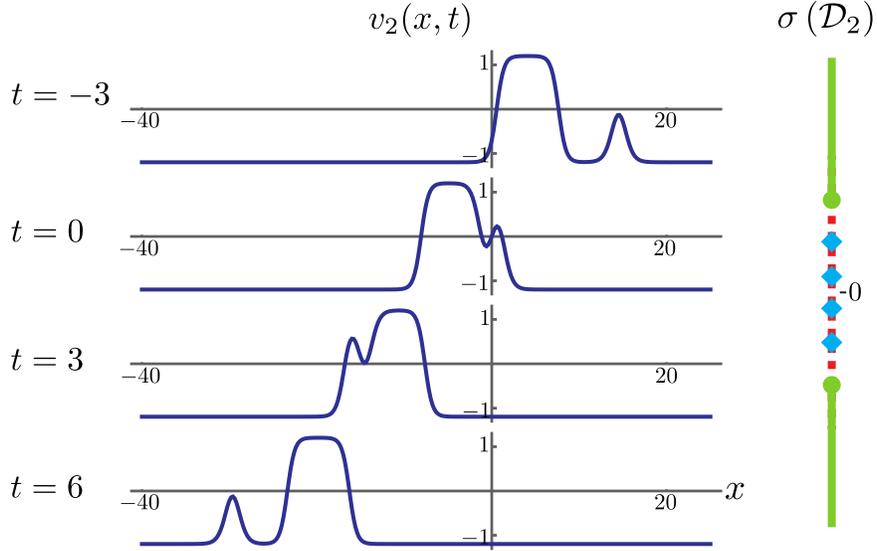}\\
	\caption{Solution of the mKdV equation 
	corresponding to two kink-antikink perturbations propagating to the left. Here
	$\kappa_1=1$, $\kappa_2=1.2$, $\kappa_3=1.2+10^{-7}$,
	 $x_{0i}=0$, $i=1,2,3$.
	 $2\kappa_3$ corresponds  to the size of the gap in the spectrum of 
	 the Dirac Hamiltonian operator,
	 $\kappa_2$  defines a bigger  kink-antikink soliton;
	 the closeness of its value to $\kappa_3$ prescribes this soliton to be higher and wider, 
	 and defines the energies $\pm \sqrt{\kappa_3^2-\kappa_2^2}$  of the 
	 two bound states
	 supported by this defect.  $\kappa_1$ defines parameters of the smaller kink-antikink soliton and 
	 the energies $\pm \sqrt{\kappa_3^2-\kappa_1^2}$ of the 
	 two  bound states supported by it.
	 In contrast with the KdV solutions with a 
	 free background, the defects with smaller 
	 amplitude propagate with higher speed.  
	 A nonzero asymptotic value of this type solution corresponds
	 to a constant mass term in a 
	 free massive Dirac Hamiltonian operator.}\label{figure7}
\end{figure}

The simplest generalization of (\ref{v0xt}) is 
given by
\be\label{mKdVk-ak}
 	v_1(x,t) =  \kappa_1 \tanh \left(\kappa_1(x+\nu t) \right)
	-\kappa_1 \tanh \left(\kappa_1(x+\nu t)-\Delta \right)
 	-\kappa_1 \coth \Delta \,,
 \ee
 where $\nu=6\kappa_2^2-4 \kappa_1^2>0$, 
 $\Delta=\frac{1}{2} 
	\log \left(\frac{\kappa_2+\kappa_1}{\kappa_2-\kappa_1}\right)>0$,
and 	$\kappa_2>\kappa_1>0$.
The mKdV solution (\ref{mKdVk-ak})  is related to the 
mutually displaced  one-soliton KdV solutions 
$u_1(x,t)$ and $u_2(x,t)$ via the Miura transformation,
$u_1=v^2_1+v'_1=\kappa_2^2-2 \kappa_1^2 
 \sech^2\left(\kappa_1(x+\nu t) \right)$,  
 $u_2=v^2_1-v'_1=\kappa_2^2-2 \kappa_1^2 
 \sech^2\left(\kappa_1(x+\nu t) -\Delta \right)$.
Analytic form of this type of the solutions is given by
relations of the same form (\ref{vmVm6k}), (\ref{VmOmOm})
but with $f(X^-_m)$ changed here for $f(X^-_m)=\exp X^-_m$.
Such solutions can be obtained from 
the topologically nontrivial solutions (\ref{VmOmOm}) 
by taking there a limit $x_{0m}\rightarrow +\infty$ or $-\infty$,
that corresponds to sending the soliton (kink)  associated 
with the non-degenerate zero energy in the spectrum of
$\cD$ to $+\infty$ or $-\infty$. 
Thus, the velocities of the remaining kink-antikink 
solitons with $j=1,\ldots, m-1$ are the same,
$\cV_j=4\kappa_j^2-6\kappa_m^2$,  as 
in the solutions with the kink, indexed there by $j=m$, and 
its traces are still present here   by 
restricting the amplitudes of the kink-antikinks  
and their speeds.
An example of such type of solutions  is shown  in Fig. \ref{figure7}.

\subsection{Kink-antikink pulse type defects in 
a
crystalline 
kink background}\label{mKdVsub3}

\begin{figure}[h]
	\centering
	\includegraphics[scale=3]{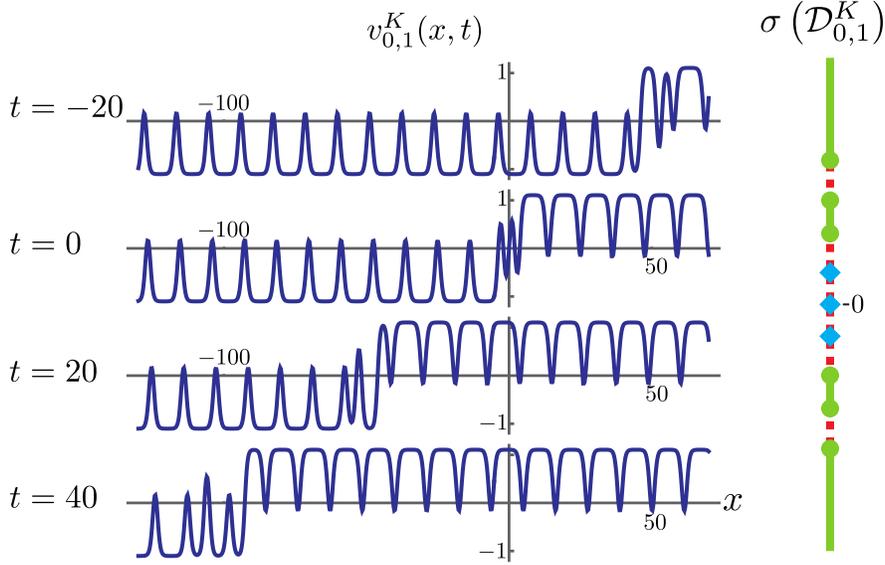}\\
	\caption{A kink-antikink (given by $\beta_1^-$) and a kink 
	(given by $\beta_2^-$) 
	type  defects propagating over a kink-antikink 
	crystalline background.
	This mKdV solution with 
	$\mu=1$, $k=0.999$, $\beta_1^-=1.7$, $\beta_2^-=1.5$ and  $C^-_i=1$,
	$i=1,2$,  
	is a somewhat analogous to that presented in Figure \ref{figure6}.
	The kink perturbation and  the kink-antikink  
	in the form of pulse (pul) soliton defect
	as well as the 
	kink-antikink crystalline background (bg)  move to the left,
	and  magnitudes of their velocities (speeds) 
	 are subject here to the relation
	$0<\vert \mathcal{V}_{kink}\vert<\vert \mathcal{V}_{pul}\vert<
	\vert\mathcal{V}_{bg}\vert$, see Section \ref{mKdVveloSec} below; 
	the pulse  soliton defect flips when passes from one to another side of   
	the crystalline kink defect.
	Notice that in contrast to the present case, 
	for the mKdV solution depicted in Figure \ref{figure6} the
	background is given there by constant asymptotes for which 
	a state of motion  is not defined. 
	The size of a central gap in the spectrum of the 
	associated Dirac Hamiltonian operator
	shown on the right is given by 
	$2\mu |\dn(\beta^-_2+i{\rm {\bf K}}'\vert k)|$.  The energies 
	of the two bound states trapped by the kink-antikink defect
	are given by 
	 $E_\pm(\alpha)=\pm \mu\sqrt{\dn^2(\alpha\vert k)-
	\dn^2(\beta^-_2+i{\rm {\bf K}}'\vert k)}$ 
	with  $\alpha=\beta^-_1+i {\rm {\bf K}}'$,
	while the zero energy value, $E_\pm(\alpha=\beta^-_2+i{\rm {\bf K}}'))=0$,
	corresponds to a unique  bound state trapped by the kink defect.
	The edges of the allowed bands 
	correspond to  $\alpha=0,\,{\rm {\bf K}},\, {\rm {\bf K}}+i{\rm {\bf K}}'$.
	}\label{figure8}
\end{figure}

On the basis of the  stationary cnoidal solution for the KdV 
equation  from Section  \ref{cnostat},
one can construct diverse types of solutions for the mKdV equation, some of which are 
topologically trivial, while others have a nontrivial topological nature.
If we use as $u_{m-1}$ a solution $u_{m-1,0}$  of the KdV equation that 
contains  only the potential well soliton defects in the periodic background,
and take  $u_m=u_{m,0}$ of the same type but with one pulse soliton defect more, 
then the associated solution $v_m$ of the  mKdV equation will describe 
multiple kink-antikink perturbations propagating in a moving crystal kink, see Figure \ref{figure8}.
The kink here is the defect of the greatest amplitude.  Together 
with other soliton defects, it propagates to the left,
see Section \ref{mKdVveloSec} below.
In this case the  solutions can be presented in the form 
\be\label{mKdVkink1}
	v^K_{0,m-1}(x,t)=V^K_{0,m}
	(x-6\lambda(\beta_{m}^-+i{\rm{\bf K}}')t,t)\,,
\ee
where
\be\label{mKdVkink2}
	V^K_{0,m}(x,t)
	=\Omega^K_{0,m-1}(x,t)-
	\Omega^K_{0,m}(x,t)\,,\qquad
	\Omega^K_{0,m}=	-
	(\log W(\mathcal{F}_+(1),\mathcal{F}_-(2),...,
	\mathcal{F}_{(-1)^{m+1}}(m)))_x\,,
\ee
and  $\lambda(\alpha)=
\mu^2(\dn^2\left(\alpha|k\right)-\frac{1}{3}(1+k'{}^2))$\,. 
The spectrum of $\cD$  is symmetric, with two finite,
 and two semi-infinite allowed  bands.
It contains a finite number of bound states in central gap, and one of these bound states 
is exactly in the center ($E=0$), while two other gaps are unoccupied.

\subsection{Multi-kink-antikink pulse type defects   
in 
a
 kink-antikink crystal background}\label{mKdVsub4}

Let us take  a solution $u_{m-1,0}$ of the KdV equation as a solution $u_{m-1}$ 
in (\ref{vmumum-1}), and choose $u_m$ in the form of the solution of the same type,
$u_m=\tilde{u}_{m-1,0}$,   but displaced by means of the 
Darboux transformation.
Then  we get the associated  solution $v_m$ of the mKdV equation in the form of 
multi-kink-antikink defects propagating in 
a
 crystalline  background. 
In this case, again, both the crystalline background as well as 
the pulse defects will propagate to the left. 
Such type of   solutions can be presented in the analytical  form
\begin{figure}[h]
	\centering
	\includegraphics[scale=3]{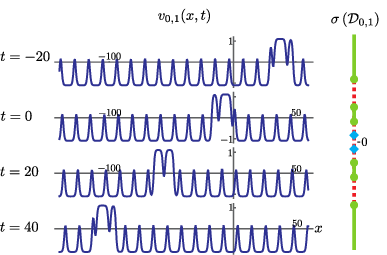}\\
	\caption{A kink-antikink pulse (given by $\beta_1^-$) 
	over a kink-antikink crystalline background. Here 
	 $\mu=1$, $k=0.9999$, $\beta_1^-=1.5+10^{-12}$, $\beta_2^-=1.5$, 
	 and $C^-_i=1$, $i=1,2$. 
	 When  the values of $\beta_1^-$ and  $\beta_2^-$ are closer, 
	 the kink-antikink defect 
	 in the background of the kink-antikink crystal is more 
	 notable. Both the kink-antikink pulse and 
	 the kink-antikink crystal background move to the left,
	 and the speed of the latter is higher than that of the defect.
	 The size of a central gap in the spectrum of the
	 associated Dirac Hamiltonian operator shown on the right
	 is given by  $2\mu|\dn(\beta^-_2+i{\rm {\bf K}}'\vert k)|$. 
	 The energies of the bound states 
	 trapped by the defect  are 
	 $E_\pm(\alpha)=\pm \mu\sqrt{\dn^2(\alpha\vert k)-\dn^2
	 (\beta^-_2+i{\rm {\bf K}}'\vert k)}$ with 
	 $\alpha=\beta^-_1+i{\rm {\bf K}}'$, and they are 
	 represented  by the blue
	 rectangles.  The non-degenerate energy 
	  values of the edges of the allowed bands  correspond  to 
	 $\alpha=0,\,{\rm {\bf K}},\, {\rm {\bf K}}+i{\rm {\bf K}}'$.}\label{figure9}
\end{figure}
\be
	v_{0,m-1}=V_{0,m}(x-
	6\lambda(\beta_{m}^-+i{\rm{\bf K}}')t,t)\,,
\ee
where 
\bea
	V_{0,m}(x,t)
	&=&
	(\log W(\mathcal{F}_+(1),\mathcal{F}_-(2),...,
	\mathcal{F}_{(-1)^{m}}(m-1)))_x\nonumber
	\\
	&&-(\log W(\mathcal{F}_+(1),\mathcal{F}_-(2),...,
	\mathcal{F}_{(-1)^{m}}(m-1),
	F(x,t,\beta^-_{m})))_x\,.\label{sinkink}
\eea
Function $F(x,t,\beta^-_{m})$ is defined here by Eq. (\ref{Fxtb}). 
The spectrum of $\cD$  is symmetric, 
with two finite and two semi-infinite allowed bands.
It has finite even number of the bound states in the central gap, and so,
in contrast with  the previous case, 
there is no bound state of discrete zero energy  in the center  
of the gap.
As in the class of solutions  discussed in the previous Section, 
the symmetric non-central gaps 
are empty of bound states. 
The mKdV solutions of this type can be obtained from those
described in the preceding Section by sending 
the kink to plus o minus infinity; the concrete analytic 
form (\ref{sinkink}) corresponds to taking 
the limit $C^-_m\rightarrow \infty$ in the solution 
(\ref{mKdVkink1}).
This case is illustrated in   Fig. \ref{figure9}.

\subsection{Multi-kink-antikink  compression  modulation defects 
over a kink in a kink-antikink crystal background}\label{mKdVsub5}

There are no nonsingular solutions of the mKdV equation associated with the 
KdV solutions which in the Darboux-Crum construction use only the states 
from the gap. To get nonsingular solutions, one can employ in the construction of $u_m$  
a state from the forbidden lower band of Lam\'e system in addition to the 
states from the gap employed for the construction of $u_{m-1}$, or 
to generate a nonlinear displacement  by means of the Darboux transformation.
In the first case we obtain a solution for the mKdV equation which contains a kink in a background, while
in the second case there will be no such kink in the structure of the mKdV solution.
The solution of the first indicated case takes 
the
 form
\be
	v^K_{2l,0}=V^K_{2l,1}(x-6\lambda(\beta_1^-+i{\rm{\bf K}}')t,t)\,,
\ee
where 
\bea
	V^K_{2l,1}(x,t)
	&=&
	(\log W(\Phi_+(1),\Phi_-(2),...,\Phi_+(2l-1),\Phi_-(2l),\mathcal{F}_+(1)))_x\nonumber
	\\
	&&-
	(\log W(\Phi_+(1),\Phi_-(2),...,\Phi_+(2l-1),\Phi_-(2l)))_x
	\,.\label{VK2l,1(x,t)}
\eea
\begin{figure}[h]
	\centering
	\includegraphics[scale=3]{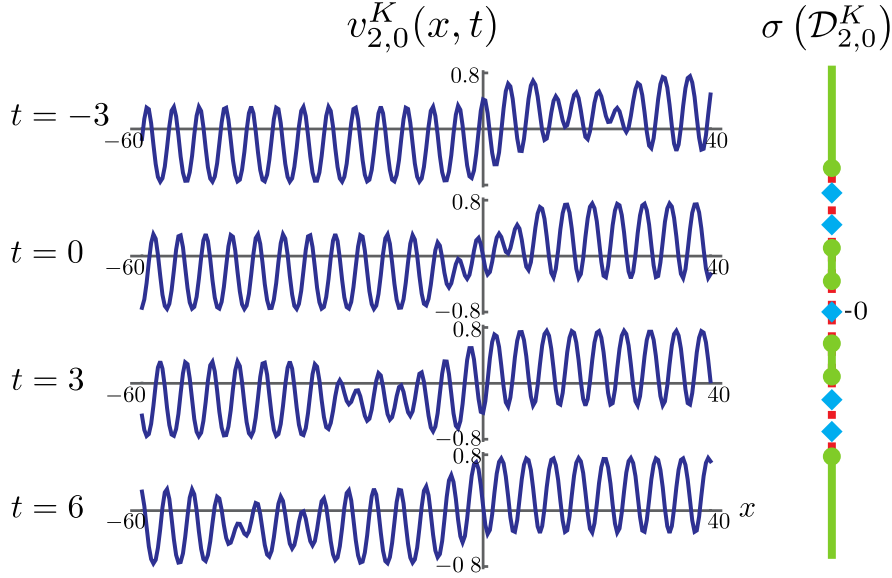}\\
	\caption{The mKdV solution with two  kink-antikink modulations  
	(given by $\beta_1^+$ and $\beta_2^+$)  
	and  a kink  (given by $\beta_1^-$ and corresponding to the 
	KdV pulse soliton) propagating  
	in a kink-antikink crystal background.
	Here
	$\mu=1$, $k=0.9$, $\beta_1^-=1.8$, $\beta_1^+=1$, 
	$\beta_2^+=1.3$ and $C^-_1=C^+_1=C^+_2=1$, and 
	the velocity magnitudes of the solitons of the compression modulations (mod) type, 
	of the background (bg), and the kink soliton are subject 
	to inequalities $\vert\cV_{mod}\vert> \vert\cV_{bg}\vert
	>\vert\cV_{kink}\vert>0$, see Section \ref{mKdVveloSec} 
	below.
	The size of the central gap in the spectrum of the associated 
	Dirac Hamiltonian operator, shown on the right,  is equal to
	 $2\mu|\dn(\beta^-_1+i{\rm {\bf K}}'\vert k)|$.
	 The energies of the bound states trapped by the kink-antikink 
	 modulations are given by 
	 $E_\pm(\alpha)=\pm \mu\sqrt{\dn^2(\alpha\vert k)-
	 \dn^2(\beta^-_1+i{\rm {\bf K}}'\vert k)}$ with $\alpha=\beta^+_1,\,
	 \beta^+_2$, while the bound state of zero energy ($\alpha=\beta^-_1+i{\rm {\bf K}}'$)
	 is trapped by the kink defect. The edges of the allowed bands in the 
	 spectrum correspond to $\alpha=0,\,{\rm {\bf K}},\, {\rm {\bf K}}+i{\rm {\bf K}}'$. 
	 }\label{figure10}
\end{figure}
In this solution, the defects of the compression modulation type propagate over 
a 
kink defect, which, in turn, propagates over the  the crystalline  background by deforming it. 
An  example of such solution is shown in Fig.  \ref{figure10}.
Here, the kink, the background, and  the compression modulation type defects
move to the left. 
The shown solution possesses an even number of compression modulation defects.
The solutions with odd number of modulation defects 
can be obtained by applying the procedures explained earlier for the KdV equation.
The spectrum of 
$\cD^K_{2l,0}$
 is symmetric, 
with two finite allowed bands and two semi-infinite allowed bands.
Besides, it has finite number of bound states in the external gaps, 
and one bound state of zero energy in the central 
gap.  

\subsection{Multi-kink-antikink compression modulation 
solutions in a kink-antikink crystal background}\label{mKdVsub6}

If the difference between 
$u_m$ and $u_{m-1}$ is a nonlinear displacement generated by 
a
 Darboux transformation, 
and if these solutions of the KdV equation 
are of the compression modulation
type defects, then the associated solution of the mKdV equation will be 
a
  kink-antikink crystal background propagating 
to the left, in which we have 
kink-antikink defects of the modulation type
also moving to the left.
Analytically such solutions  are given  by
\begin{figure}[h]
	\centering
	\includegraphics[scale=3]{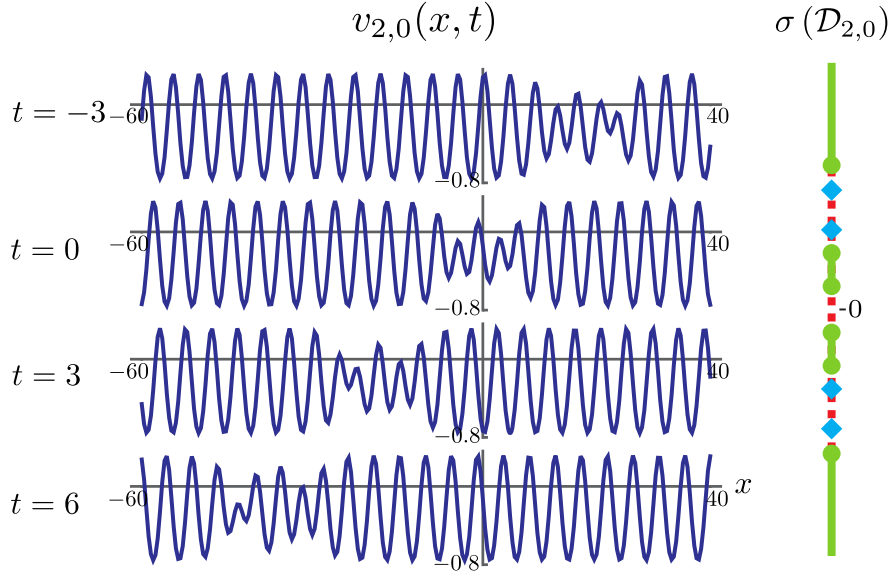}\\
	\caption{The mKdV solution with 
	two kink-antikink modulations  (given by $\beta_1^+$ and $\beta_2^+$)  
	propagating over  a kink-antikink crystal background (described by the parameter 
	$\beta^-_1$). Here $\mu=1$, $k=0.9$, $\beta_1^-=1.8$, 
	$\beta_1^+=1$, $\beta_2^+=1.3$, $C^+_{1,2}=1$. At  $t=0$, 
	the vertical axis ($x=0$) is shifted a little bit to the left
	with respect to the symmetry axis of the graphic.  This is due to the 
	limit $C_1^-\rightarrow\infty$ applied  to the solution shown on Figure \ref{figure10}
	 the present  solution. This limit corresponds to  sending  the 
	kink  to infinity and generates the  displacements (phase shifts)  
	in the background and in the remaining   defects.
	The size of the central gap in the spectrum of the associated Dirac Hamiltonian operator 
	is equal to $2\mu|\dn(\beta^-_1+i{\rm {\bf K}}'\vert k)|$. 
	The energies of the 
	bound states trapped by the kink-antikink modulation defects 
	are given by  $E_\pm(\alpha)=\pm \mu\sqrt{\dn^2
	(\alpha\vert k)-\dn^2(\beta^-_1+i{\rm {\bf K}}'\vert k)}$ with $\alpha=\beta^+_1,\,\beta^+_2$.
	 The edges of the allowed bands correspond to 
	 $\alpha=0,\,{\rm {\bf K}},\, {\rm {\bf K}}+i{\rm {\bf K}}'$. 
	 The compression defects are more rapid 
	 than the background. }\label{figure11}
\end{figure}	
\be
	v_{2l,0}=V_{2l,1}(x-6\lambda(\beta_1^-+i{\rm{\bf K}}')t,t)\,,
\ee
where
\bea
	V_{2l,1}(x,t)
	&=&
	(\log W(\Phi_+(1),\Phi_-(2),...,\Phi_+(2l-1),\Phi_-(2l),F(x,t,\beta^-_1)))_x 
	\nonumber\\
	&&-
	(\log W(\Phi_+(1),\Phi_-(2),...,\Phi_+(2l-1),\Phi_-(2l)))_x
	\,.\label{V2l,1(x,t)}
\eea
Notice a difference in the last argument of the Wronskian in the first term 
on the right in (\ref{V2l,1(x,t)}) in comparison with  (\ref{VK2l,1(x,t)}).
It reflects the fact that  the solutions of the present type 
can be obtained from those discussed in the preceding Section
by sending the kink to infinity, by taking the limit 
$C^-_1\rightarrow \infty$ in (\ref{VK2l,1(x,t)}). 

In Figure \ref{figure11}, it is shown the case with even 
number of the compression modulations  defects.
The case with the odd number of such defects can be obtained by the procedures explained 
earlier for the KdV system. 
The spectrum of 
$\cD_{2l,0}$
 is symmetric, with 
 a
 finite number of bound states appearing in non-central
gaps, and with no bound states in  the central gap.

\subsection{Multi kink-antikink modulation solitons in  
a
 kink crystal}\label{mKdVkinkcrys}

\begin{figure}[h]
	\centering
	\includegraphics[scale=3]{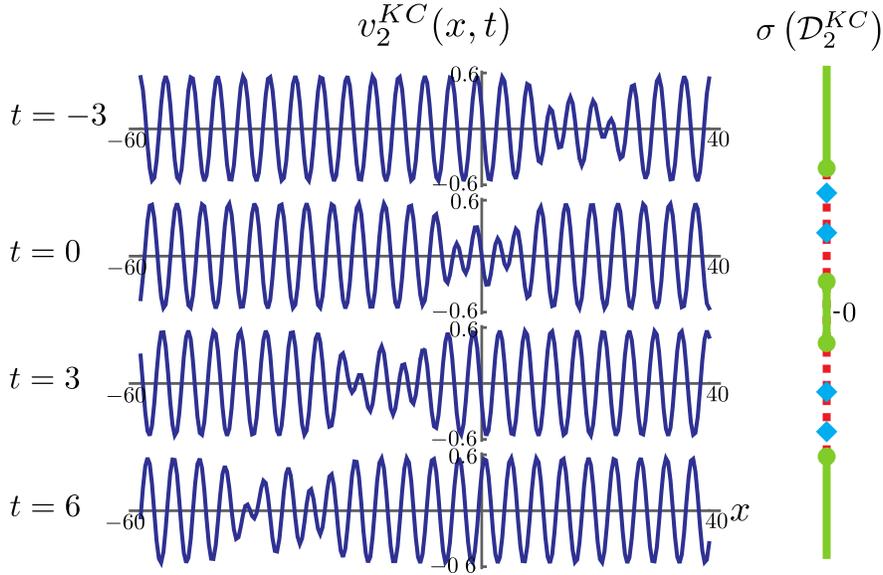}\\
	\caption{The mKdV solution with 
	two kink-antikink modulations  (given by $\beta_1^+$ and $\beta_2^+$)  
	over a kink crystal. Here,
	$\mu=1$, $k=0.9$, $\beta_1^+=1$, $\beta_2^+=1.3$, $\beta^-_1={\rm {\bf K}}$,
	$C^-_1=C^+_{1,2}=1$.
	In this case there is no central gap in the spectrum of the associated 
	Dirac Hamiltonian operator. The energies of the bound states 
	trapped by the modulation defects are given by 
	$E_\pm(\alpha)=\pm \mu\dn(\alpha\vert k)$ with $\alpha=\beta^+_1,\,\beta^+_2$, 
	while the values $\alpha=0,\,{\rm {\bf K}}$ correspond to the edges
	of the allowed bands.  In the kink crystal background, only the modulation type kink-antikink  
	defects 	can exist, and they propagate to the left more rapidly than 
	the kink crystal background. In this configuration, the velocity magnitude of the background,
	$\vert \cV_{bg}\vert =2 \mu ^2 \left(1+k'{}^2\right)$,   is minimal in 
	comparison with that in other types of the   mKdV solutions. 
		}\label{figure12}
\end{figure}

There is a special case which can be  obtained  as a limit  from the solutions discussed in the preceding 
Section.
It is generated by the choice of the solution $u_{m-1}$ of the KdV  equation 
which contains  only compression modulation defects, while the additional state 
$\Psi\propto\text{dn}(\mu x|k)$
employed for the construction of $u_m$ is at the edge of the lower forbidden band 
of the associated one-gap Lam\'e system~\footnote{This is the ground state of the Lam\'e system
at the lower edge of its valence band.}. 
In this case the crystalline background in the mKdV solution 
is centered (vertically) in zero in contrast with  the previously 
considered cases where it was displaced 
up or down. This centered crystalline background is known as the kink crystal solution 
that appears in the Gross-Neveu model \cite{Th1,PAN}. 
In comparison with the previous cases,
 here in the spectrum of $\cD$ the central gap (together with bound 
 states there) disappears and two finite continuous bands 
 merge into one central allowed  band centered at zero. 
 So, in this case one can have defects only
 of the compression modulation type, see Fig.  \ref{figure12}. 
 Here, as in the previous cases,
 the indicated soliton defects move to the left like   
the kink crystal propagating with the velocity 
 $6\lambda({\rm{\bf K}}+i{\rm{\bf K}}')$, see Section \ref{mKdVveloSec}
 below. Analytic form of such type mKdV solutions is given by
 \be
	v^{KC}_{2l}=V^{KC}_{2l}(x-6\lambda({\rm{\bf K}}+i{\rm{\bf K}}')t,t)\,,
\ee
where 
\bea
	V^{KC}_{2l}(x,t)
	&=&
	\big(\log W(\Phi_+(1),\Phi_-(2),...,\Phi_+(2l-1),\Phi_-(2l),{\rm dn}\,(\mu x\vert k))\big)_x\nonumber\\
	&&-
	\big(\log W(\Phi_+(1),\Phi_-(2),...,\Phi_+(2l-1),\Phi_-(2l))\big)_x
	\,.\label{VKC2lxt}
\eea
In the simplest case $l=0$  here the solution 
is just the kink crystal 
\be
	v^{KC}_{0}(x,t)=\big(\log {\rm dn}\,(\mu (x-\cV_{bg}t)\vert k)\big)_x
\ee
propagating to the left with velocity
 $\cV_{bg}=-2 \mu ^2 \left(1+k'{}^2\right)<0$.

\subsection{Mixed multi-kink-antikink 
solitons over a kink in 
a
kink-antikink crystal
background}\label{mKdVsub8}

In a  more general case, one can have both types  of defects, compression modulations as 
well as pulse  solitons, propagating 
over 
a
 moving crystal background. 
In dependence on whether the difference between $u_{m-1}$  and  $u_m$ 
solutions of the KdV equation 
 is a pulse type soliton or a nonlinear displacement, there will appear or not a kink defect 
in the  kink-antikink crystal
 background. In the case when the indicated difference is  a 
pulse type defect,
the solutions over the kink crystalline background take the form  
\be
	v^K_{2l,m-1}=V^K_{2l,m}(x-6\lambda(\beta_m^-+i{\rm{\bf K}}')t,t)\,,
\ee
where 
\bea
	V^K_{2l,m}(x,t)&=&
	(\log W(\Phi_+(1),\Phi_-(2),...,\Phi_-(2l),\mathcal{F}_+(1),
	\mathcal{F}_-(2),...,\mathcal{F}_{(-1)^{m+1}}(m)))_x
	\nonumber\\
	&&-
	(\log W(\Phi_+(1),\Phi_-(2),...,\Phi_-(2l),\mathcal{F}_+(1),
	\mathcal{F}_-(2),...,\mathcal{F}_{(-1)^{m}}(m-1)))_x\,
\eea
is a modulated kink.
In this solution, the kink, the pulse and modulation type defects, 
as well as  the crystalline background move 
to the left. 

\begin{figure}[h]
	\centering
	\includegraphics[scale=3]{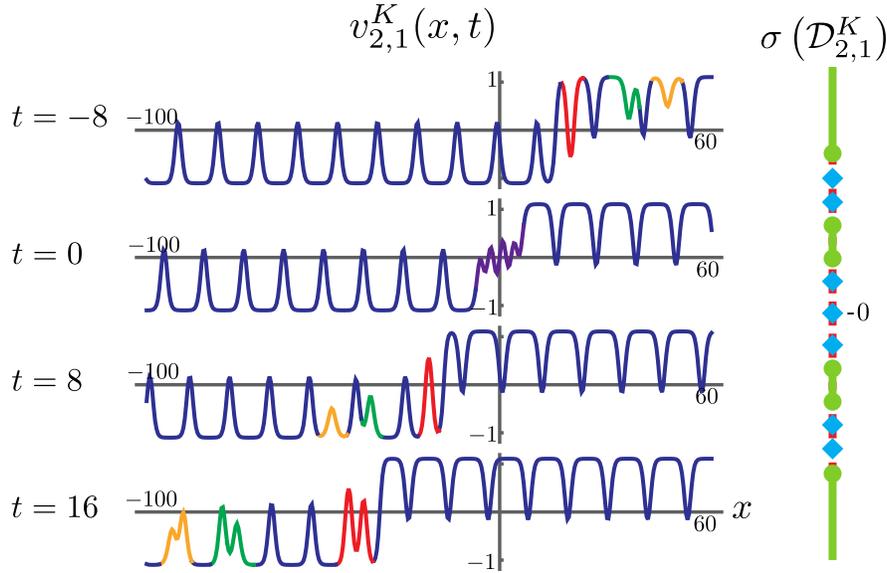}\\
	\caption{The mKdV solution with 
	a
	 kink-antikink pulse  (given by $\beta_1^-$
	and highlighted in red) 
	and two  kink-antikink  modulations (given by $\beta_1^+$ and $\beta_2^+$, and 
	shown in orange and green, respectively) propagating  over a 
	kink  defect  ($\beta_2^-$), with all 
	these
	 defects propagating in 
	a
	 moving  
	kink-antikink crystal background. 
	Here $\mu=1$, $k=0.999$, $\beta_1^-=1.7$, $\beta_2^-=1.5$, $\beta_1^+=1$, 
	$\beta_2^+=1.3$,  and  $C^\pm_{1,2}=1$, 
	and the magnitudes of velocities of the defects and background 
	are subject to inequalities 
	 $\vert \cV_{\rm mod}\vert>
	 \vert\cV_{\rm bg}\vert>
	 \vert\cV_{\rm pul}\vert>
	 \vert\cV_{\rm kink}\vert>0$. 
	 The size of the central gap in the spectrum of the 
	associated Dirac Hamiltonian operator
	 is $2\mu |\dn(\beta^-_2+i{\rm {\bf K}}'\vert k)|$. 
	 The energies 
	of the bound states trapped by the modulation defects
	are given by 
	 $E_\pm(\alpha)=\pm \mu\sqrt{\dn^2(\alpha\vert k)-
	\dn^2(\beta^-_2+i{\rm {\bf K}}'\vert k)}$ 
	with   $\alpha=\beta^+_1$, $\beta^+_2$. 
	 The energies in the central gap, which are  marked  by two  blue squares 
	 symmetric with respect to zero level,
	 are given by  
	$\alpha=\beta^-_1+i {\rm {\bf K}}'$  and correspond to the 
	bound states trapped by the kink-antikink pulse;  
	the zero energy value, $E_\pm(\alpha=\beta^-_2+i{\rm {\bf K}}'))=0$,
	corresponds to a unique  bound state trapped by the kink defect. 
	The edges of the allowed bands 
	correspond to  $\alpha=0,\,{\rm {\bf K}},\, {\rm {\bf K}}+i{\rm {\bf K}}'$.
	 }\label{figure13}
\end{figure}

The shown in Fig. \ref{figure13} solution possesses even number of compression modulation defects.
Again, solutions  with odd number of defects of this type can be obtained by means of any of  
the procedures discussed in Section \ref{case ii)}.  
The spectrum of the 
Dirac Hamiltonian operator
$\cD$ in this case is symmetric.  Besides the continuous bands 
shown on the figure,  it contains a finite number of bound states in external and central 
gaps. Besides, 
there is an additional   bound state of zero energy  in the center of the central gap,
which is  associated with  the crystalline kink.

\subsection{Mixed multi-kink-antikink defects  in 
a 
kink-antikink crystal  background}\label{mKdVsub9}

\begin{figure}[h]
	\centering
	\includegraphics[scale=3]{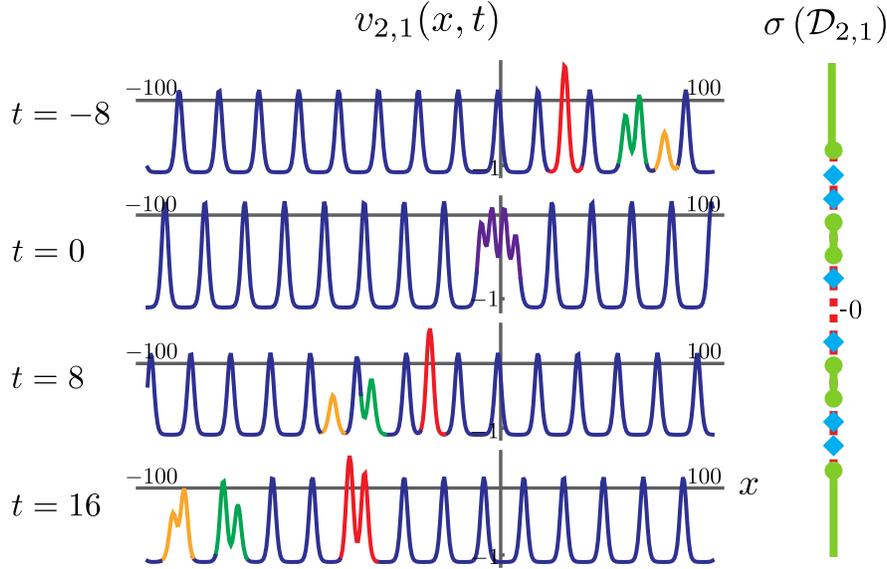}\\
	\caption{
	 The mKdV solution with 
	 a
	  kink-antikink pulse (given by $\beta_1^-$ and shown in red) 
	and two  kink-antikink modulations 
	 (given by $\beta_1^+$ and $\beta_2^+$, and highlighted in 
	 orange and green, respectively) propagating to the left in 
	 a
	  moving 
	  to the left kink-antikink crystal background. Here $\mu=1$, 
	 $k=0.999$, $\beta_1^-=1.5+10^{-10}$, $\beta_2^-=1.5$, $\beta_1^+=1$, 
	 $\beta_2^+=1.3$, $C^\pm_{1,2}=1$, and 
	 the velocity magnitudes
	  are subject to  inequalities
	 $\vert \cV_{\rm mod}\vert >
	 \vert \cV_{\rm bg}\vert >
	 \vert \cV_{\rm pul}\vert >0$.  
	 The energies in the spectrum of the associated Dirac Hamiltonian operator
	 are given by expressions similar to those in 
	 Figure \ref{figure9}. }\label{figure14}
\end{figure}

If, unlike the previous case, the difference between  $u_{m-1}$  and  $u_m$ is given by 
a nonlinear displacement, there is no kink structure in the corresponding mKdV solution
given by  
\be
	v_{2l,m-1}=V_{2l,m}(x-6\lambda(\beta_m^-+i{\rm{\bf K}}')t,t)\,,
\ee
where 
\bea
	V_{2l,m}(x,t)&=&
	(\log W(\Phi_+(1),\Phi_-(2),...,\Phi_-(2l),\mathcal{F}_+(1),
	\mathcal{F}_-(2),...,\mathcal{F}_{(-1)^{m}}(m-1),
	F(x,t,\beta^-_m)))_x
	\nonumber \\
	&&-
	(\log W(\Phi_+(1),\Phi_-(2),...,
	\Phi_-(2l),\mathcal{F}_+(1),\mathcal{F}_-(2),...,\mathcal{F}_{(-1)^{m}}(m-1)))_x
\eea
is a kink-antikink crystal
  background  displaced from zero in vertical direction
 and perturbed by propagating in it 
 mixed multi-kink-antikink defects.
The example of such a solution is depicted in Fig. \ref{figure14}.
 The symmetric spectrum of  $\cD$ has a finite number of bound states 
 in the central and external gaps, 
 with no state of zero energy in the center of the central gap.
 The mKdV solutions of this type can be obtained from those 
discussed in the preceding Section by sending the kink 
defect to infinity.
  
 \subsection{Velocities in the mKdV solutions with 
 crystalline backgrounds}\label{mKdVveloSec}
  
 The mKdV solutions $v_{m-1}$ are obtained 
on the  basis of the Darboux construction 
of KdV solutions. 
The necessary step of the procedure, as we have seen,
involves the Galilean transformation,
$x\rightarrow x-6\lambda_m t$.
The boost parameter $G=6\lambda_m$ 
of this transformation is given by the energy $\lambda_m$ 
that corresponds to the nonphysical eigenstate of the one-gap 
Lam\'e  system $L_{0,0}$ which is used to obtain the potential 
$u_m$ with one more bound state in comparison 
with the partner potential $u_{m-1}$. 
Or, this $\lambda_m$ corresponds  to the eigenstate with the help
 of which the potential $u_m$ is obtained from the completely isospectral 
 potential $u_{m-1}$  by means of the nonlinear Darboux displacement.
In both cases $\lambda_m=\lambda(\beta^-_m+i{\rm{\bf K}}')$,
and $\beta^-_m$ in the first case is associated with the 
mKdV kink supporting bound state  of non-degenerate 
discrete zero energy eigenvalue in the spectrum 
of the Dirac Hamiltonian operator $\cD$, or, if there is no 
discrete zero energy value in the spectrum of $\cD$,
 $\beta^-_m$ can be associated with the kink 
 defect sent to  infinity.
 This $\beta^-_m$ corresponds to the eigenstate of 
 the minimal energy from the lower semi-infinite forbidden band 
  in the spectrum of $L_{0,0}$ 
 which is used in the Darboux-Crum transformations
 to construct $u_m$,
 or to the eigenvalue of the  `virtual'
 state from that band associated with 
 the `kink sent to infinity'.
The crystalline background 
 in the obtained mKdV solution propagates as a 
result 
 to the left 
 and its velocity is 
 \be
	\cV_{bg}=6\lambda(\beta^-_m+i{\rm{\bf K}}')=-2 \mu ^2
	  \left(3 \text{cs}^2\left(\beta^-_m|k^2\right)+1+k'{}^2\right)<0\,.
\ee
The background propagates with a
minimal speed in  the class of the 
mKdV solutions with the kink crystal background, which 
were discussed in Section \ref{mKdVkinkcrys}.
 In that case 
 $\beta_1^-={\rm {\bf K}}$ corresponds to the 
 edge-state $\dn\,(\mu x\vert k)$ of $L_{0,0}$,
 and, therefore,    
 $\cV_{bg}=-2 \mu ^2 \left(1+k'{}^2\right)<0$. 

If the mKdV solution has the pulse type defects
different from kink, their velocities are given by
\be\label{mKdVpulsevel}
 	\mathcal{V}_{pul}(\beta^-_j)=\cV(\beta^-_j)+6\lambda(\beta^-_m+i{\rm {\bf K}}')<0\,,
 	 \quad{\rm {\bf K}}>\beta^-_j\geq \beta^-_m\,,\quad
	 j=1,\ldots, m-1\,,
 \ee
with $\mathcal{V}(\beta^-_j)$ given by Eq. (\ref{Vpulse}).
Analogously, the velocities of the modulation type 
defects, if they are present in the mKdV solution,
are  
 \be
 	\mathcal{V}_{mod}(\beta^+_j)=\mathcal{V}(\beta^+_j)+
	6\lambda(\beta^-_m+i{\rm {\bf K}}')<0\,,
 	\quad 0<\beta^+_j<{\rm {\bf K}}\,,
\ee
where $\cV(\beta^+_j)$ are given by Eq. (\ref{Vbeta+j}).
In the case if the mKdV solution has a kink defect,
its velocity is defined by relation
of the form  (\ref{mKdVpulsevel}) but with $j=m$, i.e.
\be
	\cV_{kink}=\cV(\beta^-_m)+
	6\lambda(\beta^-_m+i{\rm {\bf K}}')<0\,.
\ee

One can  see that in the mKdV solution,
the speeds of the background and defects, some types 
of which can be absent in the  solution, are ordered
according to  the inequalities $\vert \cV_{mod}\vert>\vert\cV_{bg}\vert>
\vert\cV_{pul}\vert>\vert\cV_{kink}\vert>0$.

\section{Discussion and outlook}\label{DiscOutlook}

We have constructed solutions to the KdV 
equation with an arbitrary number of solitons 
in a stationary asymptotically periodic background. In this case there exist two types of solitons:

\begin{itemize}
\item
 potential well defects (pulses), which propagate to the right,  
 \item 
 compression modulation defects, which move to the left.
 \end{itemize}
 
 \noindent These solutions asymptotically have a form 
 of the one-gap Lam\'e potential  
 but subjected to the  phase shifts $x\rightarrow x-x^\mp_0$, 
 $x^\mp_0=\pm \frac{1}{\mu}(\sum_i\beta^-_i+\sum_j\beta^+_j)$
 for $x\rightarrow \mp\infty$ with respect to the stationary 
 solution (\ref{u0Lame}).
The asymmetry in propagation of the two types 
of soliton defects  is valid in  the case of a stationary background. 
If we apply 
Galilean transformations to the KdV solutions, 
we obtain new solutions, for  which, in general case,
 the described  propagation asymmetry   
of the defects over the now moving background will be changed. 
However, this does not change
the picture of the relative motion: pulse defects  
always will propagate to the right with respect to the asymptotically 
periodic background, while modulation type defects
will move to the left with respect 
to non-stationary crystalline background.
The interesting peculiarity we also have observed 
in the constructed KdV solutions is that in the 
limit cases when amplitudes of the pulse and compression modulation 
defects tend to zero,
the limit values of the velocities of defects with respect to
the crystalline background are nonzero.

For the mKdV equation, we have constructed the following solutions 
from the obtained KdV solutions by means of the 
Miura-Darboux-Crum 
transformations: 

\begin{itemize}

\item
solutions with 
a  kink crystal background, 
in which there can exist only solitons in the form 
of the compression modulation type defects,

\item
 solutions with 
 a
   kink-antikink crystal background, in which 
 there can exist kink-antikinks in the form of the pulse
 and/or compression modulations type defects.
 \end{itemize}
 \noindent
  In 
  a
    kink-antikink 
 crystal background there also can appear  a topological defect 
 in the form of the  kink (or antikink, if we use a symmetry of the 
 mKdV equation by  changing 
 $v(x,t)$ for $-v(x,t)$) which always is
 related with the KdV pulse type defect
 that traps the bound state with the lowest energy 
 in the lower forbidden band in the spectrum 
 of the associated perturbed one-gap Lam\'e system.
 
Unlike the KdV, the mKdV equation has no Galilean symmetry, and 
the velocities of the defects and asymptotically periodic 
background in the solutions we constructed have an absolute
character. 

In the mKdV solutions, 
all the  defects and  crystalline backgrounds  move to the left,
and the velocity magnitudes  of the kink-antikink defects 
of the modulation (mod) and  the pulse (pul) types,
and the velocity magnitudes   of the background (bg) and 
the kink 
are subject to the inequalities 
$\vert \cV_{mod}\vert>\vert\cV_{bg}\vert>
\vert\cV_{pul}\vert>\vert\cV_{kink}\vert>0$.
Thus, with taking into account  the sign of
the velocities, we have for the defects and backgrounds in the 
mKdV solutions, 
similarly to those in  the KdV solutions, $(\cV_{pul}-\cV_{bg})>0$
and $(\cV_{mod}-\cV_{bg})<0$.
At the same time, the velocity magnitude (speed)  
of the kink defect, if it is present
in the mKdV solution, always has a  minimal value in comparison 
with other velocity magnitudes.

The presence or absence of the kink in the mKdV solution
is detected by  $N=2$ supersymmetry of
 the associated extended Schr\"odinger system.
It  is  generated by the 
 first order supercharge  operators $\cS_a$, $a=1,2$, 
having a nature of the Dirac Hamiltonian 
operators with a scalar potential.
When kink is present or absent, the supercharges possess 
or not a zero mode, and 
the
$N=2$ supersymmetry is unbroken or broken.
The supersymmetry detects also 
the case of the mKdV solutions with the  kink crystal background.
For such solutions, the kernel of supercharges is 
two-dimensional.
Unlike the solutions with kink, the corresponding zero modes 
in this case  are given by not normalizable but 
periodic  states~\footnote{
These zero modes are constructed from 
Darboux-dressed edge states 
$\dn (\mu x\vert k)$  and 
$\dn (\mu x+{\rm {\bf K}}\vert k)$
of the pair of mutually shifted in the 
half-period Lam\'e systems by applying to them  
the Galilean boost with velocity $\cV_{bg}=-2\mu^2(1+k'{}^2)$.}.
The position of the bound states in gaps
in the spectra of the supercharges 
defines also the  type of 
defects 
present in the corresponding mKdV solution.

We have showed that the  $N=2$ supersymmetry
constitutes a part of a more broad, $N=4$ type exotic 
nonlinear supersymmetry,
which includes in its structure two bosonic generators 
composed  from the nontrivial, Lax-Novikov integrals of the 
pair of the Schr\"odinger subsystems. These bosonic generators  
are higher derivative 
differential operators, one of which is the central element 
of the superalgebra. Besides, exotic supersymmetric structure
contains 
additional pair of supercharges being matrix differential operators
of the even order. The additional 
integrals appropriately  reflect 
the peculiar nature of the extended
Schr\"odinger system associated with the pair of the KdV solutions
by detecting all the bound states and the band-edge states 
in the spectrum, as well as distinguish the eigenstates 
 corresponding to the fourth-fold 
degenerate  energy values inside  the allowed bands.

The both,  KdV and mKdV equations are invariant under simultaneous 
 inversion of $t$ and $x$ variables, but not under reflection of $t$ 
 (or, of $x$) only.
The stationary trivial solution $u=0$, on the basis
 of which we construct soliton solutions over the asymptotically free
 background, is invariant under
 separate inversions in $t$ and $x$. 
 The same is true for the stationary 
 cnoidal solution 
 in the form of the stationary 
one-gap Lam\'e potential.
 So,  the `initial data' in the construction are invariant under $t$-inversions,
 and anisotropy of evolution as well as 
the  chiral asymmetry of the non-stationary  solutions 
 is rooted  in the anisotropy 
 of  the equations themselves. 
 Namely, in our construction,
 though  the `seed' KdV solutions are 
 time-inversion invariant,
the solutions of the auxiliary problem 
in the Lax formulation for the KdV equation,
which are generating elements for the Darboux-Crum transformations,
 break this symmetry. 
 In the case of the asymptotically 
 free background, in the indicated 
 solutions of the auxiliary problem 
 the dependence on time 
 enters universally  in the form of  arguments
 $x-\cV_j t$ with  $\cV_j=4\kappa^2_j>0$.
 As a result, all the solitons in non-singular 
 KdV solutions 
move to the right. 
For  the soliton solutions over  
an
 asymptotically 
 periodic background, the third order Lax operator
 of the initial Lam\'e system
 has eigenvalues of different signs 
 on the upper and lower 
 horizontal borders of the 
 fundamental $\alpha$-rectangle
which correspond to 
 the lower forbidden band and the gap.
 It is this sign asymmetry  that finally is responsible for
 chiral asymmetry in propagation of the KdV solutions 
 in the form of pulse and compression modulation
 defects over 
 a
  crystalline background.
The asymmetry of the mKdV solutions is inherited from that
for the KdV solutions.

 \vskip0.1cm 
 
 Since  the KdV and mKdV solutions, and particularly those associated
 with the Lam\'e quantum system,  find many diverse applications 
 in a variety of different areas of physics 
 ranging  from hydrodynamics, 
 plasma physics, and optics
 to hadron physics and cosmology 
 \cite{Lamb,Kivshar,Th1,BD1,Th2,GrayWater,BDT1,Lam4,
 Lam5,Lam6,Lam7,Lam8,Lam9}, it would 
 be very interesting to find 
 where the obtained new solutions could show up themselves.
 They could appear in the form 
 of perturbations of different nature,   
 which would  propagate in a nonlinear media 
 with different velocities and reveal chiral asymmetry 
 in their dynamics. Another peculiarity which 
 could be associated with the described solutions 
 is the existence of nonzero  bounds for the 
 velocity of the defects with disappearing amplitudes. 
 
\vskip0.1cm 

It would also be interesting to
consider a generalization of  the  approach
employed  here 
by using Darboux transformations 
for the 
first order Hamiltonian operator  of a (1+1)-dimensional Dirac 
system instead of the second order Schr\"odinger operator. 
In this way one could  get 
finite-gap 
Dirac Hamiltonian operators 
of the form  
\be
	\cD=\left(\begin{array}{cc}V_2(x)& 
	-\frac{d}{dx}+V_1(x)\\ \frac{d}{dx}+V_1(x)& -V_2(x)\end{array}\right)
\ee
with asymmetric spectrum. 
The corresponding  stationary potentials  $V_{1,2}(x)$
could then be promoted  to  solutions 
in the form of the twisted kinks  and twisted kink-antikinks
\cite{BD1,BDT1,CDP,CorJak} in a periodic 
background for nonlinear Schr\"odinger equation 
belonging to the  Zakharov-Shabat -- Ablowitz-Kaup-Newell-Segur
 hierarchy.

In  the supersymmetric quantum mechanical  structure we discussed,
the $N=4$  refers   to the number of supercharges
appearing in the extended Schr\"odinger system.
As the extended system is composed from  a pair of
the perturbed one-gap Lam\'e systems,
one could expect the appearance of only 
two supercharges as it happens in supersymmetric
quantum mechanical systems of a general nature
\cite{CoKhSu}.
The peculiarity of the considered systems consists 
in their finite-gap nature, and it is this
property that is behind the extension of the 
usual $N=2$ 
supersymmetry up to the exotic
$N=4$   nonlinear supersymmetric 
structure that incorporates the pair of Lax-Novikov
integrals of the subsystems in the form of the  two additional 
bosonic generators.
It would be interesting to investigate  if
the described exotic supersymmetric
structure can be related somehow to the supersymmetric 
extensions of the KdV and mKdV equations 
and corresponding hierarchies that are 
considered within the superspace (superfield)
generalizations  of the indicated classical 
$(1+1)$-dimensional integrable systems
\cite{Kupersh,Mathieu,ChaiKul,BeIvKr,IvKrTo,Top}.
 
 \vskip0.3cm

\noindent \textbf{Acknowledgements.}
We are grateful to Francisco  Correa and Francesco Toppan 
for useful comments.
The work has been partially supported by FONDECYT Grant
No. 1130017.   A. A. also acknowledges 
the  CONICYT  scholarship 21120826 
and financial support of Direcci\'on de Postgrado and 
Vicerrectoria Acad\'emica of the Universidad de
Santiago de Chile.


\end{document}